\documentclass[iop]{emulateapj}

\usepackage{amssymb}
\usepackage{multirow}
\usepackage{morefloats}
\usepackage{amsmath}
\usepackage{bigstrut}
\usepackage{booktabs}
\usepackage{natbib}
\usepackage{float}
\usepackage{graphicx,epsfig,fancyhdr,epsf,txfonts,epstopdf}
\usepackage{latexsym,bbm}
\usepackage{lineno}
\usepackage{color}
\usepackage{ulem}

\slugcomment{Not to appear in Nonlearned J., 45.}

\shorttitle{Flares in the blazar S5 0716+714}
\shortauthors{Xu et al.}

\begin{document}

\title{Statistical analysis of micro-variability properties \\ of the blazar S5 0716+714}

\author{Jingran Xu\altaffilmark{1,$\dagger$}, Shaoming Hu\altaffilmark{1,2,$\star$},James R. Webb\altaffilmark{2}, Gopal Bhatta\altaffilmark{3},Yunguo Jiang\altaffilmark{1}, \\ Xu Chen\altaffilmark{1}, Sofya Alexeeva\altaffilmark{1} and Yutong Li\altaffilmark{1}}
\affil{Shandong Provincial Key Laboratory of Optical Astronomy and Solar-Terrestrial Environment, \\ Institute of Space Sciences, Shandong University, Weihai, 264209, China\\}
\email{$\dagger$ xujingrichard@foxmail.com}
\email{$\star$ corresponding author: husm@sdu.edu.cn}

\affiliation{$^{2}$Florida International University, 11200 SW 8th St, Miami, FL 33199, USA\\}

\affiliation{$^{3}$Astronomical Observatory, Jagiellonian University, 30-244 Krakow, Poland}

\begin{abstract}
  The typical blazar S5 0716$+$714 is very interesting due to its rapid and large amplitude variability and high duty cycle of micro-variability in optical band. We analyze the observations in I, R and V bands obtained with the $1.0m$ telescope at Weihai observatory of Shandong University from 2011 to 2018. The model of synchrotron radiation from turbulent cells in a jet has been proposed as a mechanism for explaining micro-variability seen in blazar light curves. Parameters such as the sizes of turbulent cells, the enhanced particle densities, and the location of the turbulent cells in the jet can be studied using this model. The model predicts a time lag between variations as observed in different frequency bands. Automatic model fitting method for micro-variability is developed, and the fitting results of our multi-frequency micro-variability observations support the model. The results show that both the amplitude and duration of flares decomposed from the micro-variability light curves confirm to the log-normal distribution. The turbulent cell size is within the range of about 5 to 55 AU, and the time lags of the micro-variability flares between the I-R and R-V bands should be several minutes. The time lags obtained from the turbulence model are consistent with the fitting statistical results, and the time lags of flares are correlated with the time lags of the whole light curve.
\end{abstract}


\section{Introduction}

 The study of active galactic nuclei (AGN) is an important field in exploring the physical mechanism of extra-galactic astronomy. As a special subclass of the AGN, the blazar class is composed of BL Lacertae (BL Lacs) and flat spectrum radio quasars (FSRQs). According to the AGN unified model, the jet direction of blazar is close to our line of sight \citep{Urr95}. Blazars are characterized by high and variable polarization, synchrotron emission from relativistic jets and spectral variability in all wavelengths ranging from radio to $\gamma-$ray \citep{Wag95, Bot03}. The continuum spectrum radiation of the blazar is generally considered to be dominated by non-thermal radiation. Spectral energy distributions (SED) are in the shape of a double peaked pattern. The low energy component has a peak within IR-to-X-ray range and is usually attributed to Doppler-boosted synchrotron radiation \citep{Bre81, Urr82, Mar98, Bla79}. The high energy component peak in the MeV-TeV energy range is likely produced by inverse-Compton (IC) process \citep{Mar92}. Regarding the source of seed photons, there are two possible contributions to radiation from inverse-Compton process: synchrotron self-Compton (SSC) \citep{Ghi93, Tak96, Tav98, Kin02} and external-radiation-Compton (ERC) \citep{Ghi96}. Large amplitude and rapid variability is an evident property of AGN. The variability time scales are often diverse and usually range from several minutes to years.

 This rapid and large amplitude variability of blazars provides a unique opportunity to study the internal physical processes of the jet, such as particle acceleration, relativistic beaming effects, origin of flares and size, structure, and location of the emission regions. The physical mechanism of different spatial scales can be explored by studying the light curves in different time scales. The variability time scales of blazars can be roughly divided into three categories: the variation over few minutes to less than a day known as intra-day variability (IDV), also known as micro-variability \citep{Mil89, How04, Aga15}; the short-term variability (STV) having the timescales from days to months; the variation on a timescale of months to years called the long-term variability (LTV). Flares may be caused by inhomogeneous collisions in shock waves that are propagating along the jet at relativistic velocities \citep{Bla00}. The observation of flares in micro-variability provides an opportunity to study particle injection and acceleration at the sub-parsec scale of the jet. In this process, the kinetic energy and internal energy of the jet are transformed into the kinetic energy of the particles and finally into radiation.

 Although some observers remain skeptical about the existence of micro-variability, many observations and analytical methods have been established for the micro-variability of AGN \citep{Men17, Car92, Vil02, Gau15, Hag02, Pol07}. \cite{Gup09} and \cite{Hon18} reported the optical quasi-period oscillation for S5 0716+714, which were explained by accretion disk fluctuations or oscillations. Very long baseline interferometry (VLBI) observations of the blazar jets in the form of radio knots proved the existence of postulated compact regions, and the apparent super-luminal motion \citep{Bar86}. It is a direct observational proof that relativistic activity exists in the jet. It implies that there may be small-scale structures in the jet and provides evidence for the existence of micro-variability.

 So far, there are many models explaining how particles interact in the jet. One of the most widely accepted models for the generation of flares for large amplitude short-term variations is the collisions between inhomogeneities moving with different velocities in the jet \citep{Sik94}. \cite{Leh89} described the spectral characteristics of X-rays through the shot model. The shot contour obtained by Lehto has poor similarity with the micro-variability observed in S5 0716$+$714. The model used in this work is the turbulent model in the plasma jet proposed by \cite{Mar92}. Following \cite{Bha13}, it is considered that the observed flares are caused by turbulence in the jet. Particles in each turbulence plasma cell are accelerated by strong shock and cooled by synchrotron radiation, resulting in flares of the light curves. The observed micro-variability light curves are the result of the convolution of all flares. By analyzing the characteristics of flares distribution and characteristics of turbulence in the jet can be studied. It is difficult to fit the model to large data sets so we developed an algorithm to fit the model to a large amount of data automatically. An automatic fitting method is proposed to solve this problem, and the possible flares in the light curve are fitted by theoretical analysis.

 Blazar S5 0716$+$714 (DA 237, HB89) with a redshift of z=0.31$\pm$0.08 is an ideal object for studying BL Lac-type objects \citep{Nil08}. This object was chosen due to its high duty cycle of micro-variability and lots of well sampled simultaneous multi-frequency observations. It is one of the blazars with largest amplitude and variability in the optical radiation region. The variation rate can be faster than 0.10$\sim$0.12$mag\cdot hr^{-1}$ \citep{Sag99, Vil00, Wu05, Mon06, Hu14a}, and its variation timescale can be as short as 15 minutes \citep{Sas08, Ran10, Cha11}. Although some investigations have discussed the time lag between different bands of the source \citep{Lia14, Wu12}, the time lag of micro-variability is not clear. Based on the observation data of Weihai Observatory of Shandong University, the time lag of the micro-variability light curve is associated with the turbulent flare model.

 In section 2, the observation details and data reduction methods of S5 0716$+$714 are introduced. The models and theoretical analysis used for flares are described in section 3. In section 4, the results of statistical analysis are shown and discussed. Section 5 is dedicated to the summary and conclusion.

\section{Observations and Data Reductions}

 The observations were obtained in I, R and V bands with the $1.0m$ telescope at Weihai observatory of Shandong University from 2011 to 2018 \citep{Hu14b}. The telescope is a classical Cassegrain design with a focal ratio of f/8, and it is equipped with a back-illuminated PIXIS 2048B CCD camera or iKon-L 936 CCD. Quasi-simultaneous optical multi-band observations provide colour behavior and time lags between different bands. So sequence of I, R and V bands images were taken repeatedly as long as possible on clear night. Standard differential photometry method was used to reduce all frames. We selected four stars from the sequence of \cite{Vil98} for comparison. Stars 1 and 2 are usually overexposed due to their high brightness and cause problems when trying to obtain accurate photometry, so they were excluded from comparison sequence. The stars 3, 5 and 6 were selected as comparison stars.

 \begin{figure*}
  \begin{minipage}{\textwidth}
  \centering
  \includegraphics[trim=0.6cm 0.5cm 0cm 1.6cm,width=0.48\textwidth,clip]{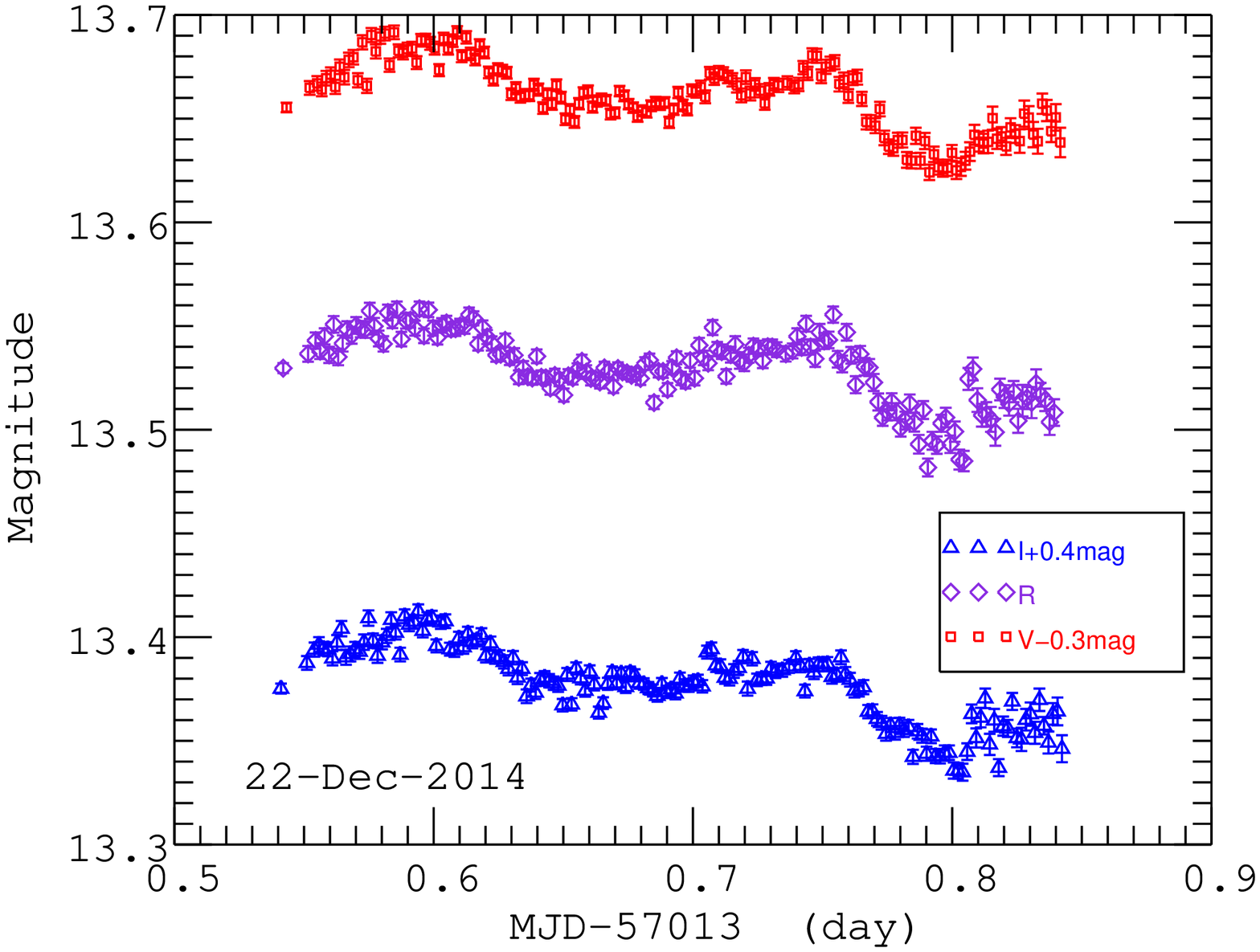}
  \includegraphics[trim=0.6cm 0.5cm 0cm 1.6cm,width=0.48\textwidth,clip]{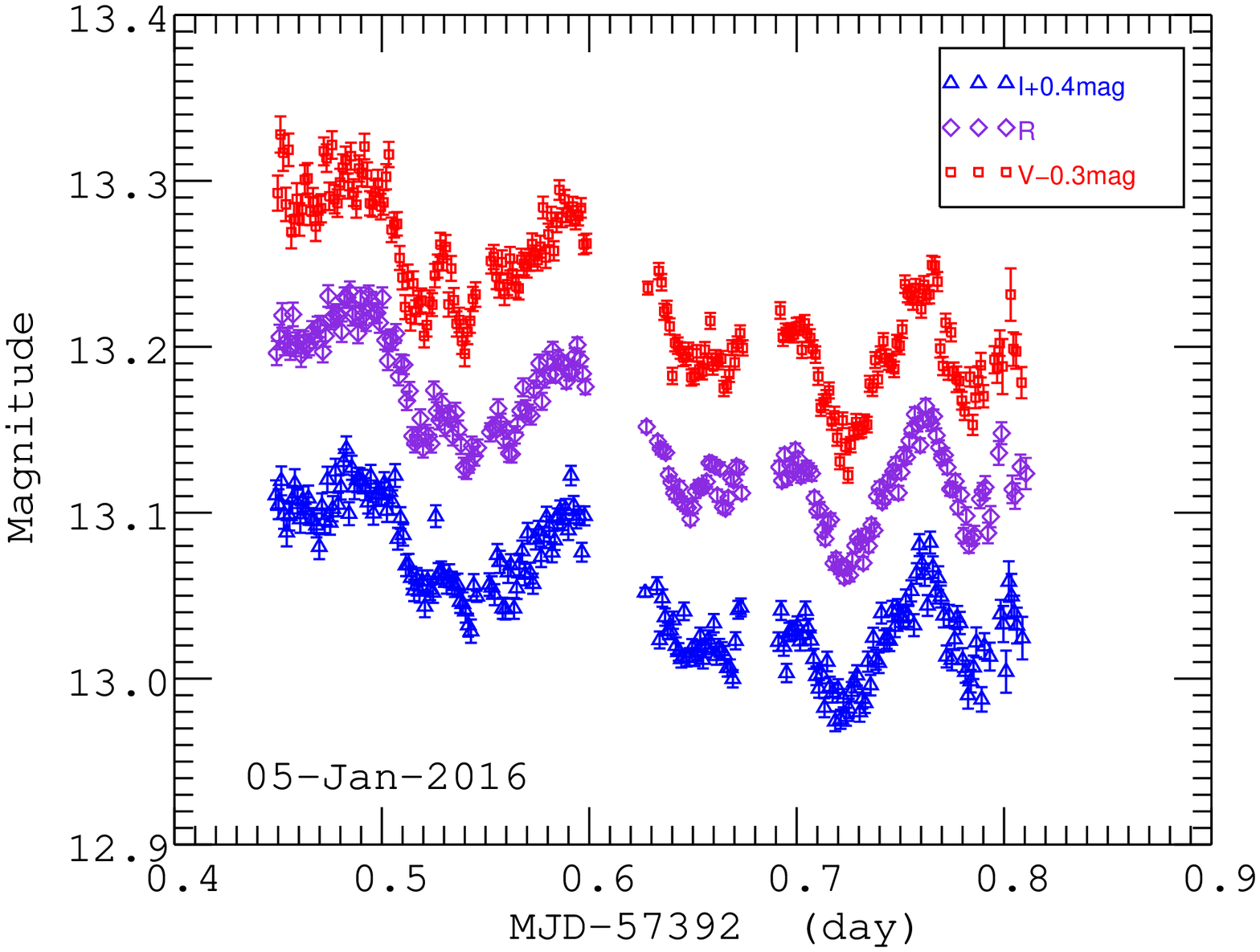}
  \includegraphics[trim=0.6cm 0.5cm 0cm 1.6cm,width=0.48\textwidth,clip]{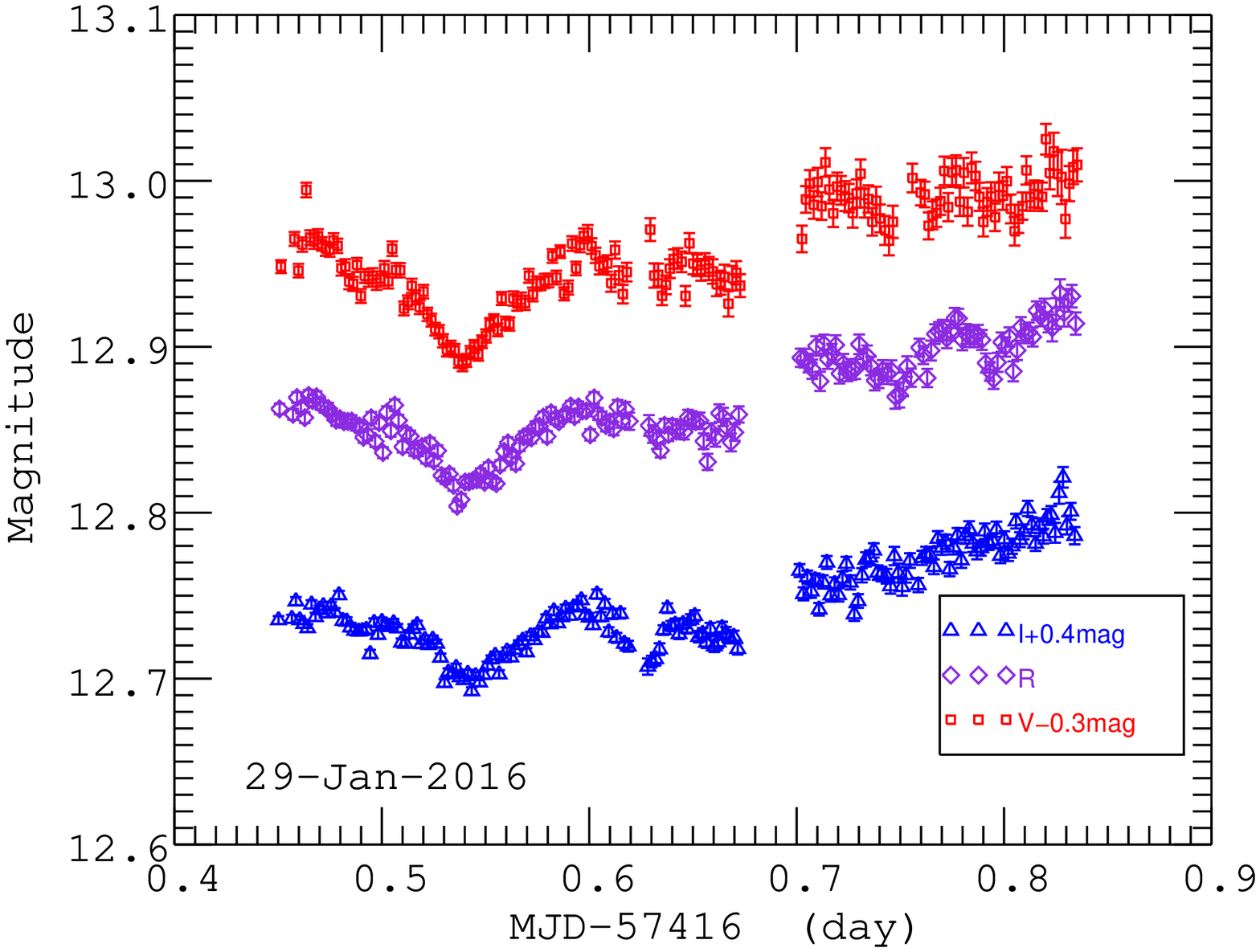}
  \includegraphics[trim=0.6cm 0.5cm 0cm 1.6cm,width=0.48\textwidth,clip]{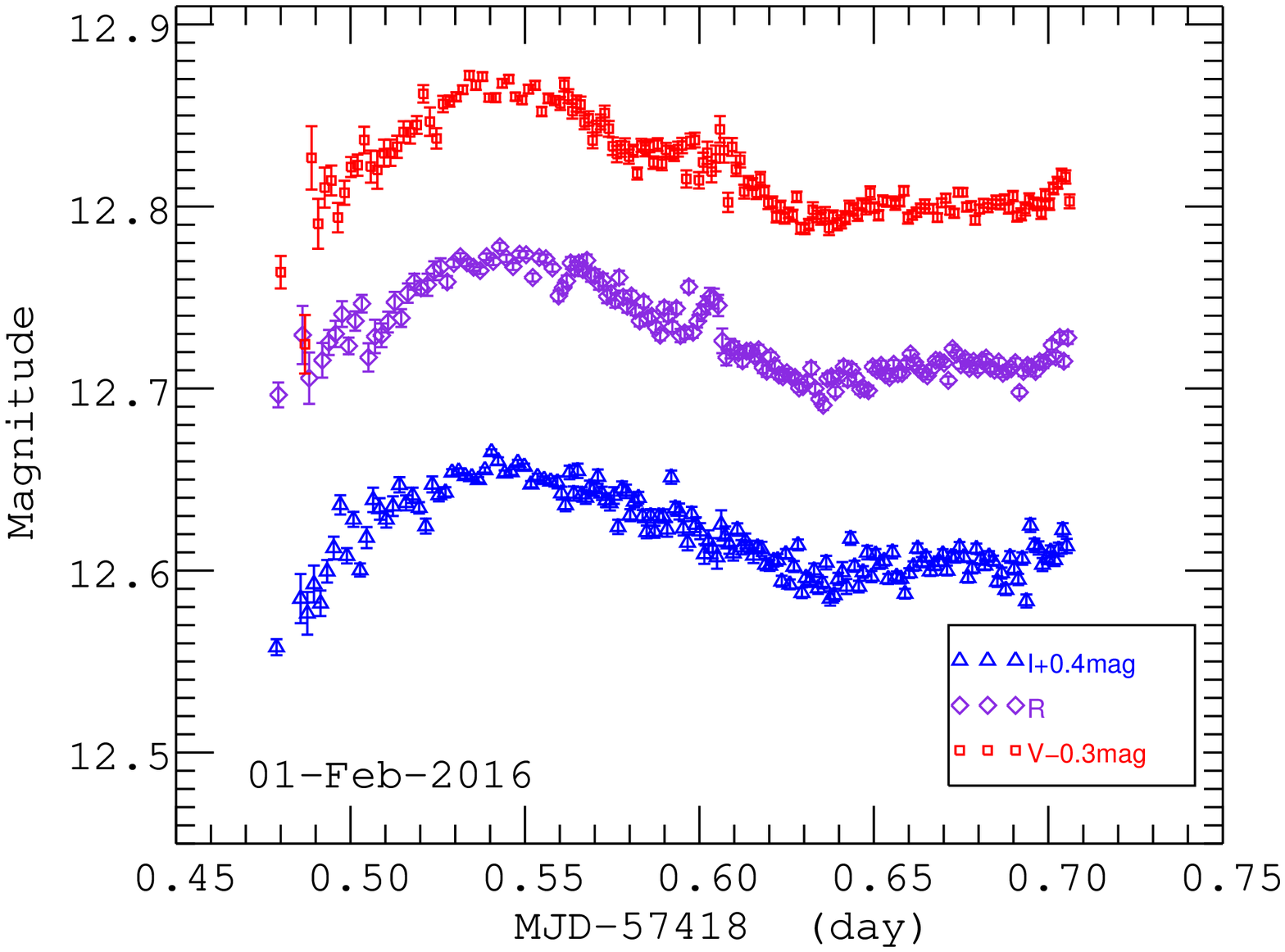}
  \end{minipage}\vspace{0.001cm}
  \caption{Four examples of IDV light curves of S5 0716+714 from Weihai observatory of Shandong University. The triangles, rhombuses and squares represent the light curve in I, R and V band, respectively. \textbf{The I, R and V band photometry shown in panel of this figure is available as the Data behind the Figure.}}
  \label{fig:lightcurve}
 \end{figure*}

 All object frames were automatically processed by the aperture photometry program compiled by \cite{Chen14}. The photometric data are divided into different IDV light curves according to the date, and select 75 light curves with high observation accuracy and variability characteristics. The total data consist of 16786 high quality observations, including 5465 data points in I band, 5664 data points in R band and 5657 data points in V band. The observation data of S5 0716$+$714 from 2011 to 2014 have been used in \cite{Hu14a}, where the IDV test and color variation of micro-variability were studied. The full versions of Table \ref{tab:phosample} are only available as online electronic form. Four IDV light curves are shown as examples in Figure \ref{fig:lightcurve}. In order to insure the accuracy of the model fits, we only use the observations with photometric error less than 0.08mag. The standard flux conversion of I, R and V filters given by \cite{Bessell13} is used to convert the magnitude into flux. 
 
 \begin{table}
 	\centering
 	\caption{Examples of photometric results catalog}
 	\label{tab:phosample}
 	\begin{tabular}{ccccc}
 		\tableline\tableline
 		Observation Date & Band & Julian Day & Magnitude & Error \\\tableline
 		20160105 & I & 2457392.949 & 12.71 & 0.01 \\ 
 		20160105 & I & 2457392.950 & 12.71 & 0.01 \\ 
 		20160105 & I & 2457392.952 & 12.72 & 0.01 \\
 		20160105 & R & 2457392.949 & 13.20 & 0.01 \\
 		20160105 & R & 2457392.951 & 13.21 & 0.01 \\
 		20160105 & R & 2457392.952 & 13.22 & 0.01 \\
 		20160105 & V & 2457392.950 & 13.59 & 0.01 \\
 		20160105 & V & 2457392.951 & 13.63 & 0.01 \\
 		20160105 & V & 2457392.952 & 13.62 & 0.01 \\\tableline\tableline
 	\end{tabular}
    \tablecomments{\textbf{Table \ref{tab:phosample} is published in its entirety in the machine-readable format. A portion is shown here for guidance regarding its form and content.}}
 \end{table}

 A smoothing algorithm is applied for data processing in order to eliminate the noise interference in very short time scales during the model fitting. The variation rate cannot be too fast according to the observational data, thus the brightness difference between adjacent data points will not be too large when the time duration between adjacent data points is small. If their difference value is large, the data points will be smoothed using the IDL smoothing algorithm. The smoothing algorithm only affects the results of the time scale in a few minutes, which is much shorter than any possible period we can find from the time domain analysis. In order to ensure that no unexpected errors are introduced into smoothing, discrete Fourier transform analysis (DFT) of all time domain ranges of Non-smoothed data is also carried out. The results are the same as those obtained from smoothed data in the time domain we are concerned with. The light curves after smoothing were compared with the data before smoothing. The smoothing process does not change the trend of the light curve, and removes some high-frequency noise in the data. In addition, the DFT method is used to analyze the frequency domain characteristics of several light curves. The results of this study show that the frequency distribution of each light curve is random.

 Although the observations include a lot of IDV data in S5 0716$+$714 optical bands, the sample is still discontinuous for some IDV light curves. In order to improve the accuracy of data analysis and obtain reliable results, all continuous light curves with no less than 10 data points in an hour were selected. Because it is impossible for an observatory to observe continuously for several days without interruption, all the selected light curves span between 1 and 10 hours. For data with slightly less continuity, i.e. data points with adjacent time intervals greater than 5 minutes but less than 30 minutes, cubic spline interpolation was used. The purpose is to reduce the error caused by the boundary value of discontinuous points in the fitting process. The fitting parameters obtained by this interpolation are not included in the statistical results, so this interpolation will not affect the final model fitting results.

 Systematic analysis of micro-variability light curves shows that the light curves can be modeled reasonably as a convolution of many pulses. These pulses should be caused by the flares of small-scale structures, so the influence of large-scale structures should be excluded. In order to avoid introducing STV or LTV conversion in the fitting, it is necessary to eliminate the STV$/$LTV trend in the light curves, that is, the light curves need to be substracted a baseline component. If the light curves are long enough, polynomials can be used to fit the minimum values of all flares to obtain the baseline flux curve. The minimum flux of the daily light curve was used as the baseline flux, because the used daily light curve does not exceed 10 hours. The residual light curves after subtracting the baseline flux will be used in the model fitting below as a superposition result of the flares.

\section{Model Analysis}

There are many investigators reported about micro-variability in blazars, e.g. \cite{Hum08, Web10, Web16, Qui91}. Many studies showed that micro-variability light curves can be resolved into pulses or shots. We used the model based on individual rapid synchrotron pulses caused by a shock encountering individual turbulent cells in the jet, developed by \cite{Bha13}. Since close inspection of many micro-variability light curves show that the light curves can be resolved into pulses or shots, they suggested a model where individual synchrotron cells are energized by a plane shock propagating down the jet and result in an increase in flux resembling a pulse. It is supposed that the superposition of these pulses is the cause of the micro-variations seen in the light curve. Compared with previous studies \citep{Abd10}, this study focuses on the properties of IDV light curves. The model in this work can combine observation data with physical mechanism together. The physical parameters of the small-scale structure of the jet in the optical radiation region can be obtained and analyzed.

\subsection{Micro-variability model}

 \begin{figure*}
    \centering
    \includegraphics [trim=0.6cm 0.3cm 0.8cm 0.4cm,width=0.8\textwidth,clip]{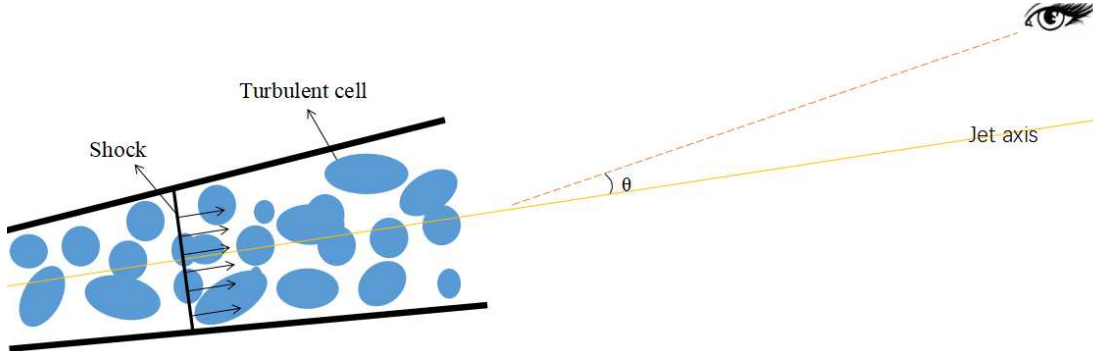}
    \caption{A schematic picture illustrating the approach adopted in solving the Kirk diffusion equations. When the shock propagates along the jet direction and passes through each turbulent cell in turn, the particles in the cell are rapidly energized and accelerated. Subsequently, after the shock passes through the cell, each energized cell loses energy due to isotropic synchrotron radiation.}
    \label{fig:schematic}
 \end{figure*}

 In this work, we used the random synchrotron radiation cells model, which is proposed by \cite{Mar92}. According to \cite{Bha13}, micro-variability is caused by turbulence prevailing in the jet. Each independent turbulent cell may have different density, size and magnetic field direction. The formation of these turbulent cells is a statistical process. By decomposing the light curve into individual flares, the basis of turbulence in the jet can be understood better. In order to obtain the physical parameters, the formation mechanism and pulse profile of the flare forming the micro-variability light curve are theoretically studied. For the pulse profile in the micro-variability light curves, we used the equation of \cite{Kir98}, hereafter KRM. The KRM equation shows that when a plane shock encounters a cylindrical density enhancement region, particles with various magnetic field directions and particle density are accelerated in front of the shock. The geometric structure of the model is shown in Figure \ref{fig:schematic}.

 In this model, it shows that the synchrotron radiation intensity will increase with the increase of particle injection rate. On the contrary, the radiation intensity decreases as the particles cool after the shock passes through the turbulent cell. The increased magnetic field intensity or particle density will lead to the increase of particle injection rate, thus the rising edge of the flare appears. After the shock propagated the turbulent cell, the particle injection rate returns to the initial value, thus the observed flare descending edge appears. The convolution of these single flares results in the micro-variability light curves. The observed micro-variability light curves are the result of the turbulent cells interacting with propagating shock waves in turn.

\subsection{Flare profile}

\begin{figure}
 	\centering
 	\includegraphics[trim=0.7cm 0.7cm 2cm 1.8cm,width=0.6\textwidth,clip]{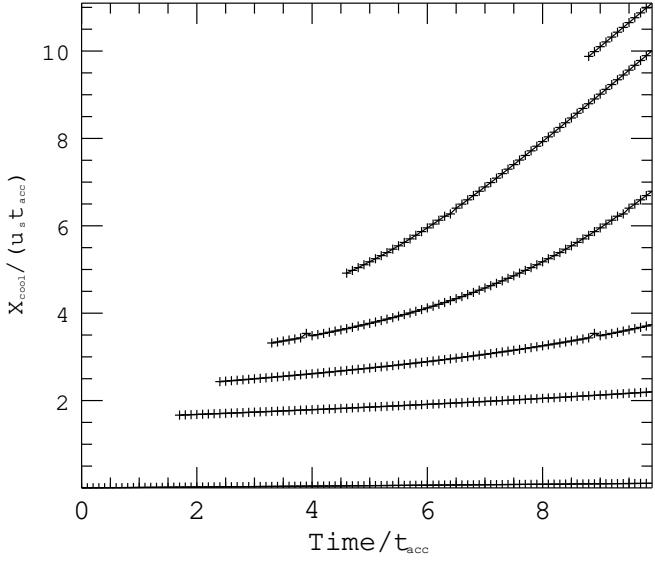}
 	\caption{Distribution of cooling length with time and Lorentz factor $\gamma$ = 10, 50, 100, 200, 500, 1000 from bottom to top, respectively.}
 	\label{fig:xcool}
 \end{figure}

 Based on the previous work, we assume that the electrons accelerated by the shock follow the KRM equations \citep{Kir94},
 \begin{equation}
 \frac{\partial N}{\partial t} + \frac{\partial}{\partial \gamma}[(\frac{\gamma}{t_{acc}} - \beta_{S}\gamma^{2})N] + \frac{N}{t_{esc}} = Q\delta (\gamma - \gamma_{0})
 \label{equ:KRM}
 \end{equation}
 where,
 \begin{equation}
 \beta_{S} = \frac{4}{3} \frac{\sigma_{T}}{m_{e}c} (\frac{B^{2}}{2\mu_{0}})
 \label{equ:betas}
 \end{equation}
 with $\sigma_{T}$ Thomson scattering cross section. $N$ is the electron number density in the energy space expressed by Lorentz factor $\gamma$. $t_{acc}$ and $t_{esc}$ are acceleration time and escape time, respectively. The second term in brackets in Eq.(\ref{equ:KRM}) describes the rate of energy loss due to synchrotron radiation averaged over pitch-angle in a magnetic field $B$ (in Tesla), where $\mu_{0}$ and $m_{e}$ are permeability of free space and mass of an electron, respectively.The particle distribution in Eq.(\ref{equ:KRM}) can be solved by Laplace transformation method, and the particle distribution in acceleration and radiation regions can be obtained,
 \begin{equation}
 N(\gamma,t) = a \frac{1}{\gamma^{2}} (\frac{1}{\gamma} - \frac{1}{\gamma_{max}})^{(t_{acc} - t_{esc})/t_{esc}} \times \Theta(\gamma - \gamma_{0})\Theta(\gamma_{1}(t) - \gamma)
 \label{equ:N}
 \end{equation}
 \begin{equation}
 \begin{split}
 n(x,\gamma,t) & =  \frac{a}{u_{S}t_{esc}\gamma^{2}} [\frac{1}{\gamma} - \beta_{S}(t - \frac{x}{u_{S}}) - \frac{1}{\gamma_{max}}]^{(t_{acc} - t_{esc})/t_{esc}} \\ & \times \Theta[\gamma_{1}(\frac{x}{u_{S}})-(\frac{1}{\gamma}-\beta_{S}t + \beta_{S}\frac{x}{u_{S}})^{-1}]
 \label{equ:n}
 \end{split}
 \end{equation}
 where $\Theta(\gamma)$ is the Heaviside step function, $ \gamma_{max} = (\beta_{S}t_{acc})^{-1}$, $u_{S}$ is shock velocity, $\gamma_{0}$ is the Lorentz factor at the initial time and $\gamma_{1}$ represent the upper limit of particle energy. And $a$ is given as,
 \begin{equation}
 a=Q_{0}t_{acc}\gamma_{0}^{t_{acc}/t_{esc}} (1-\frac{\gamma_{0}}{\gamma_{max}})^{-t_{acc}/t_{esc}}
 \label{equ:a}
 \end{equation}
 By integrating Eq.(\ref{equ:n}), the particle density can be obtained,
 \begin{equation}
 \begin{split}
 \int n(x,\gamma,\bar{t})dx = & \frac{a}{(1 - \frac{u_{S}}{c}) \gamma_{max}^{(t_{acc} + t_{t_{esc}) / t_{esc}}}} (\frac{\gamma_{max}}{\gamma})^{2} \\ \times & \Big \{ [\frac{\gamma_{max}} {\gamma} - \frac{\bar{t}}{t_{acc}} + \frac{x(1-u_{S}/c)}{u_{S}t_{acc}} - 1]^{\frac{t_{acc}}{t_{esc}}} \Big \} _{x_{0}(\gamma,\bar{t})} ^{x_{1}(\bar{t})}
 \label{equ:intn}
 \end{split}
 \end{equation}
 where $\bar{t} = t- x/c$, the value of the lower integral limit $x_{0}$ is constrained by the cooling length $x_{cool}$ and the maximum spatial range of the radiation region. The following transcendental equation can be obtained by solving the range of values in Eq.(\ref{equ:n}),
 \begin{equation}
 \frac{\gamma_{max}}{\gamma} - \frac{\bar{t} + x_{cool}/c} {t_{acc}} + \frac{x_{cool}}{u_{S}t_{acc}} - 1 = (\frac{\gamma_{max}}{\gamma_{0}} - 1)^{-\frac{x_{cool}} {u_{S}t_{acc}}}
 \label{equ:transcendental}
 \end{equation}
 Figure \ref{fig:xcool} is the numerical analysis of the solution of the transcendental Eq.(\ref{equ:transcendental}). Different curves represent the $x_{cool}$ distribution of $\gamma = 10,50,100,200,500,1000$ from bottom to top, respectively. It shows that the lower integral limit $x_{cool}$ is related to the retarded time $(\bar{t}=t-x/c)$ and Lorentz factor $(\gamma)$, and can not be expressed by simple analytical expressions. Therefore, the integral solution of Eq.(\ref{equ:transcendental}) is carried out by numerical method. From the analysis by Figure \ref{fig:xcool}, it can be concluded that there is a switch time $t_{on}$ for particles with different Lorentz factor $\gamma_{c}$, i.e., there is no particle greater than this energy in the plasma accelerated by shock when $t \leq t_{on}$. Conversely, particles with energy $\gamma_{c}$ exist only when $t \geq t_{on}$. The analysis shows that there is a range of particles with energy $\gamma_{c}$ in time domain, which corresponds to the upper and lower bounds of spatial integration. The existence range of a given energy particle corresponds to its radiation range, but it is not equal to the spatial range of turbulence.
 
\begin{figure}
 	\centering
 	\includegraphics[trim=0.7cm 0.6cm 2cm 1.8cm,width=0.6\textwidth,clip]{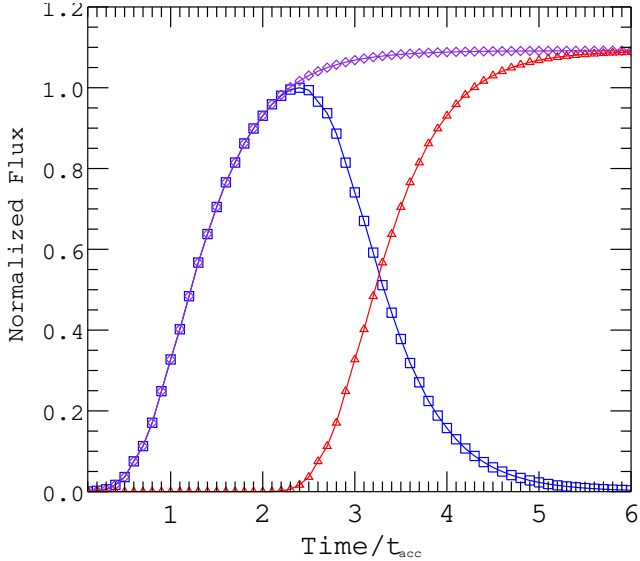}
 	\caption{The profile of flare derived from the KRM equation is shown by the square solid line in the figure. The diamond solid line is the rising curve when the injection rate of particles increases. The drop function of the particle injection rate returning to the base value after the shock wave passes through the cell is shown by the triangular solid line in the figure.}
 	\label{fig:pulse}
  \end{figure}

  After the particle distribution is obtained, the synchrotron radiation intensity is calculated by integrating the Green's synchrotron function $P(\nu,\gamma)$ for each frequency.
  \begin{equation}
  I_{0}(\nu,t) = \iint P(\nu,t)n(x,\gamma,\bar{t} + x/c)dxd\gamma
  \label{equ:i0}
  \end{equation}
  The variation of strength with time at a particular frequency is shown by the diamond solid line in Figure \ref{fig:pulse}.
 
  It is assumed that the bulk velocity of the jet material is $0.9c$ \citep{cel08}, and the shock moves at the velocity of $0.1c$ relative to the jet material \citep{Kir98}. For the generation of a single flare, it can be considered as the result of the following equation:
 \begin{equation}
 I(\nu,\bar{t}) = I_{1}(\nu,\infty) + \eta_{f}[I_{1}(\nu,\bar{t}) - I_{1}(\nu,(1 - u_{S}/c)t_{f})]
 \label{equ:itotal}
 \end{equation}
  Where $t_{f}$ is the pulse duration time. Based on the observed data, the fundamental flux $I_{0}$ is determined by the whole light curve, and it is assumed that this is the background radiation of the laminar flow in the jet. For the observed fundamental flux, the injection rate constant $Q_{0}$ can be deduced by Eq.(\ref{equ:i0}), which is the particle injection of the laminar flow. The change of particle injection rate $\triangle Q$ in each cell determines the amplitude of the flare. For the $Q$ parameter in KRM equation, it is mainly affected by the magnetic field intensity $B$ and the angle $\theta$ between the magnetic field and the line of sight. Considering the amplitude change is the result of their interaction. For flares at $0 < t < t_{f}$, $Q(t)=(1+\eta_{f})Q_{0}$ can be derived from Eq.(\ref{equ:i0}) and Eq.(\ref{equ:itotal}), where $\eta_{f}Q_{0}=\Delta Q$. The Eq.(\ref{equ:itotal}) shows that the duration time $t_{f}$ of the flare, which corresponds to the size of turbulent cells in the model. The pulse profile of the flare is determined by the value of $t_{acc}/t_{esc}$ when the duration of the flare is determined. When $t_{acc} \sim t_{esc}$, the profile of flare is approximately symmetrical, and the ratio of $t_{acc}/t_{esc} = 0.5$ is fixed here. The profile of flare is shown in Figure \ref{fig:pulse}. This profile of flare is used as the standard for automatic fitting and statistical analysis.
  
  In these equations, particle acceleration time scale and escape time scale not only restrict the length of cooling zone, but also control the pulse profile. The amplitude of flare is related not only to the enhanced particle density, but also to the magnetic field intensity and direction $\theta$. In this model, the cylindrical jet can be divided into two regions, the acceleration region and the radiation region. The density enhancement region where the shock propagating is the acceleration region. The area where the energized particles stay until they are cooled is the radiation region.

\begin{figure}
    \centering
     \includegraphics[trim=1.0cm 1.2cm 1.6cm 1.8cm,width=0.45\textwidth,clip]{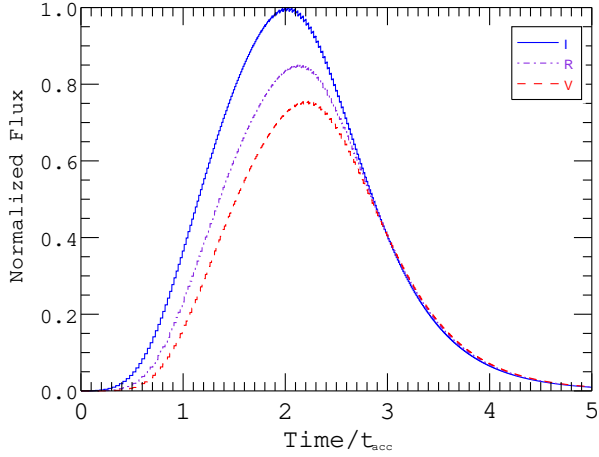}
     \caption{The flare response of different bands is calculated theoretically. Pulse shape for frequencies $\nu_{I} = 3.79 \times 10^{14}$, $\nu_{R} = 4.68 \times 10^{14}$ and $\nu_{V} = 5.45 \times 10^{14}$ from Eq.(\ref{equ:itotal}) with $B = 2$ Gauss and time of flare $t_{f} = 1.5t_{acc}$. In order to compare the profile characteristics of different frequencies conveniently, the flux in each band is normalized.}
     \label{fig:diffband}
 \end{figure}

 Based on the analysis of the effects of different physical parameters in KRM equation on the radiation region and pulse profile, the phenomena of micro-variability observed can be reasonably explained. Figure \ref{fig:diffband} shows the flare profiles obtained from KRM equation in different bands. Through this comparison and calculation, the time delay between flares at different frequencies in optical bands can be obtained. The time delays between I and R, R and V band calculated by this method are about $12.98_{-12.98}^{+9.23}$ minutes. It should be noted that the theoretical time delay is related to the profile of flare.

\begin{figure}
    \centering
     \includegraphics[trim=1.0cm 1.2cm 1.6cm 1.8cm,width=0.45\textwidth,clip]{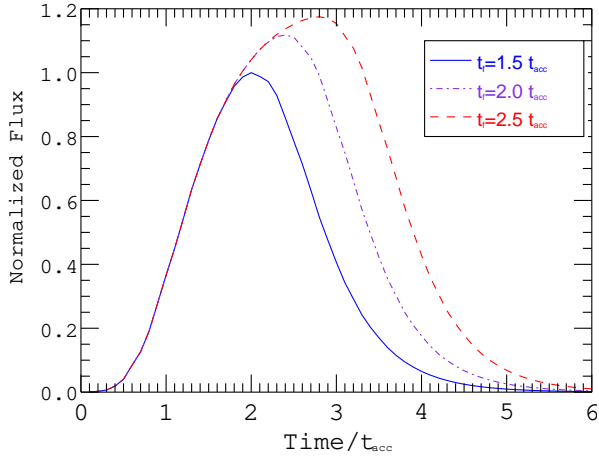}
     \caption{Three profiles of flare corresponding to different durations of flare. Solid line, dash dot line and dashed line represent the corresponding profile in different duration scales. Pulse shape for durations $t_{f} = 1.5t_{acc}$, $t_{f} = 2.0t_{acc}$ and $t_{f} = 2.5t_{acc}$ from Eq.(\ref{equ:itotal}) with $B = 2$ Gauss and optical frequency $\nu = 3.79 \times 10^{14}$. The flux in each band is normalized.}
     \label{fig:difftf}
 \end{figure}

 Figure \ref{fig:difftf} shows the profile comparison of different flare duration derived from KRM theoretical simulation. In these results, the flare profiles of different duration are different. Further calculation shows that the time lag between them has changed. The result shows that turbulent cell of different sizes may have different time lag. This case may be one of the reasons for the time delay between micro-variability light curves in different bands. It indicates that the time delay and flare duration, the proportion of particle acceleration time scale and escape time scale are related.

\subsection{Automated Fitting Routine}

 \begin{figure*}
  \begin{minipage}{\textwidth}
  \centering
  \includegraphics[trim=0.7cm 0.2cm 0cm 1.8cm,width=0.45\textwidth,clip]{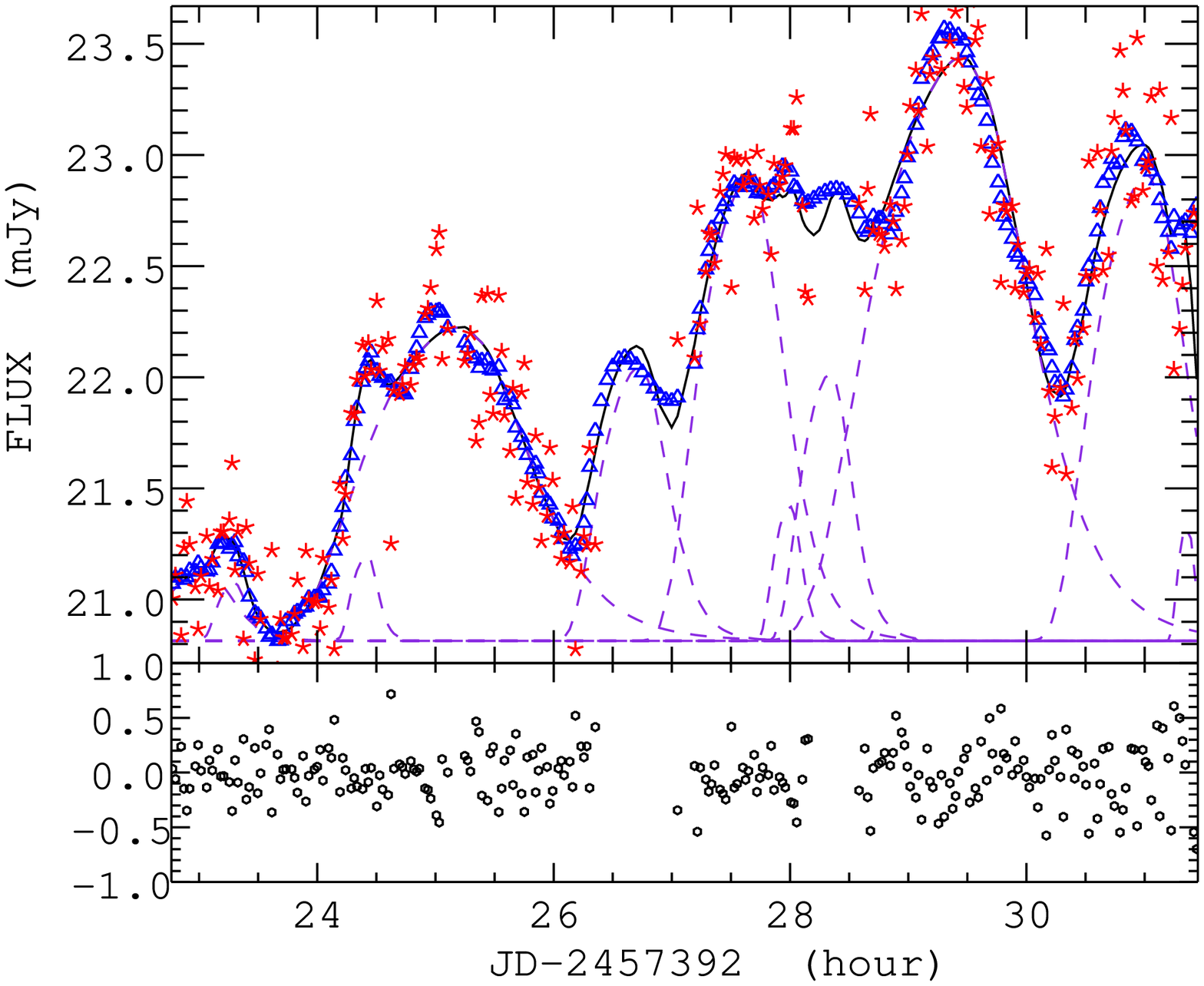}
  \includegraphics[trim=0.7cm 0.2cm 0cm 1.8cm,width=0.45\textwidth,clip]{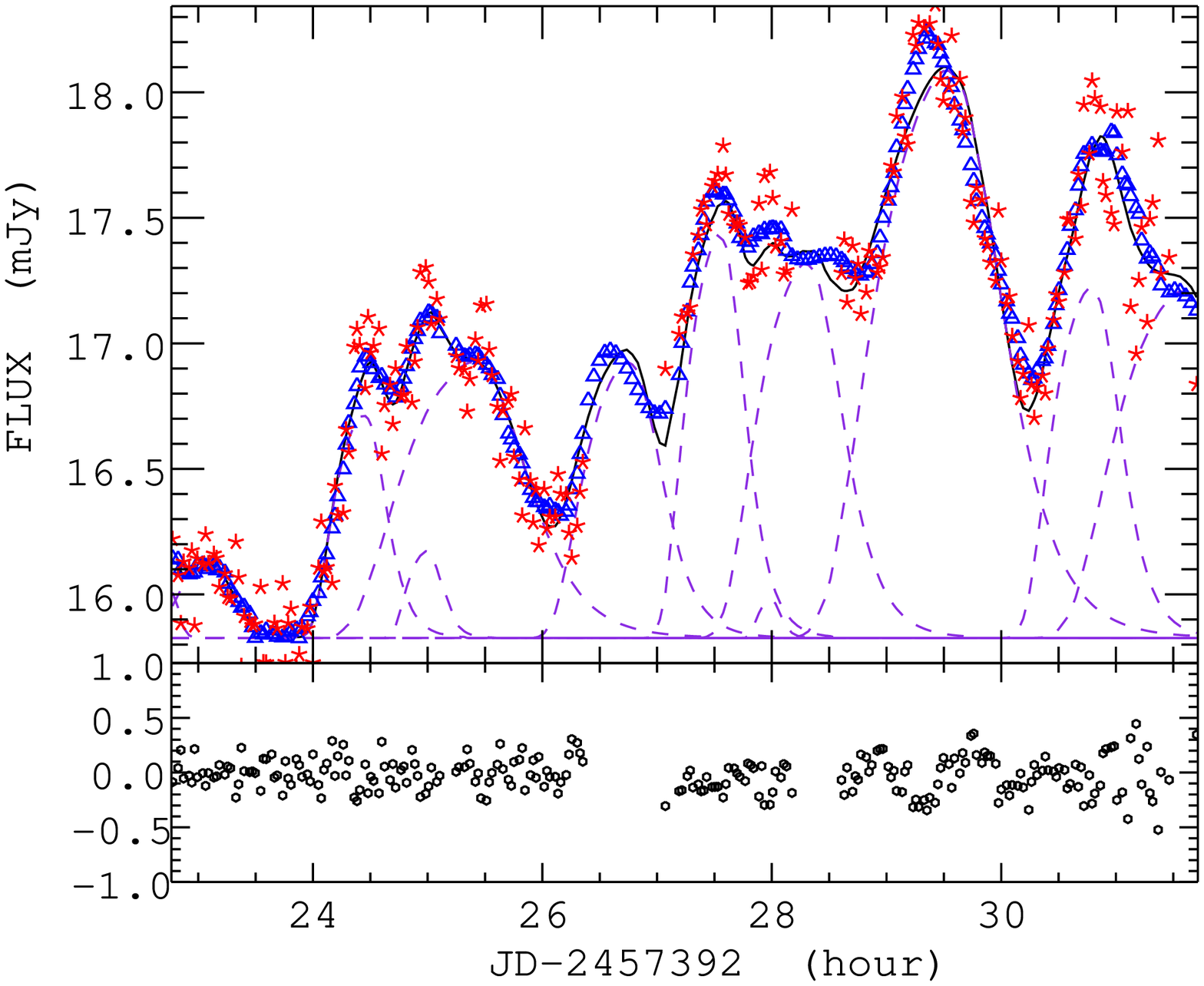}
  \includegraphics[trim=0.7cm 0.2cm 0cm 1.8cm,width=0.45\textwidth,clip]{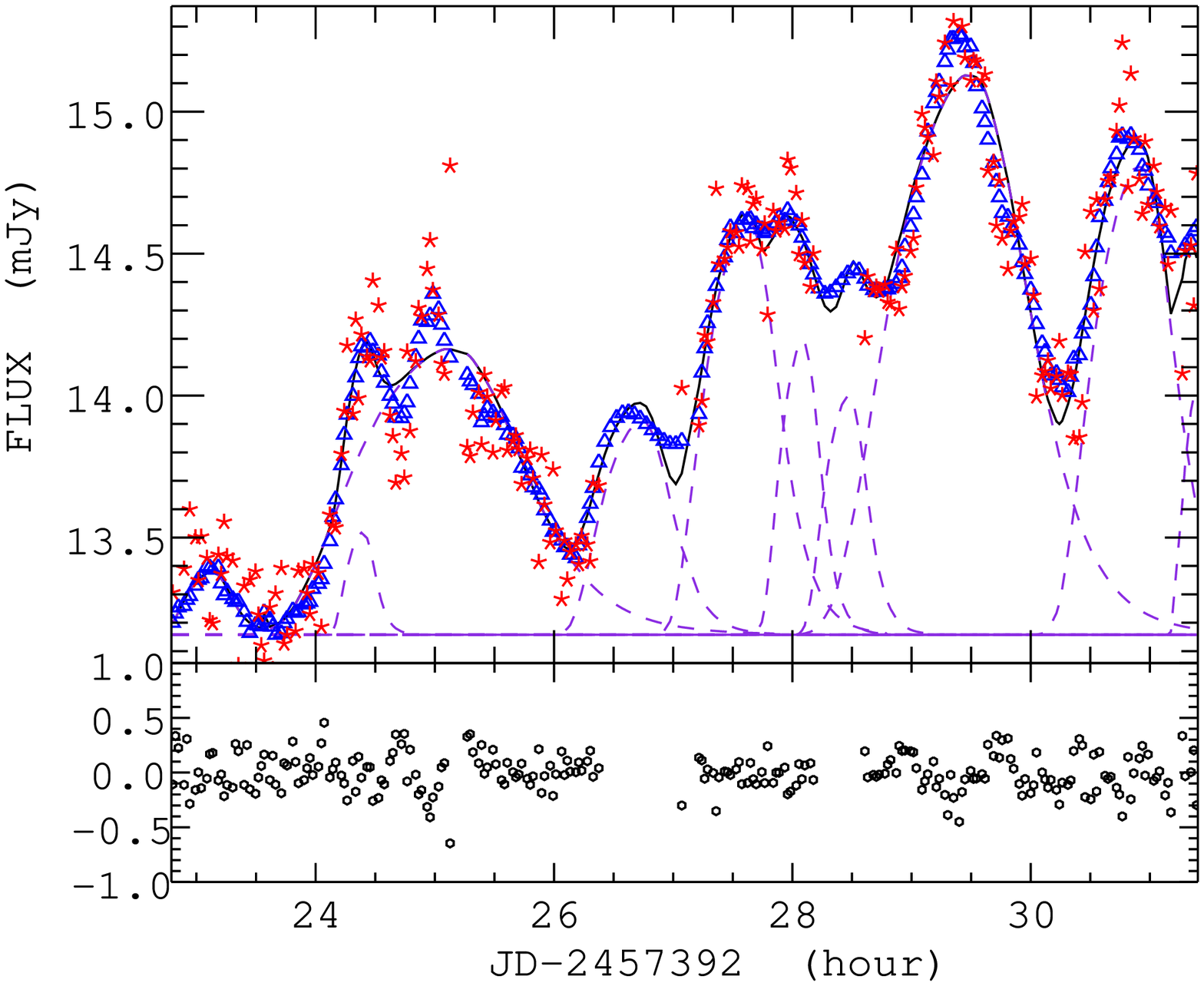}
  \end{minipage}\vspace{0.001cm}
  \caption{Examples for decomposition of continuous micro-variability light curves of S5 0716$+$714 into individual flares in I (left), R (right) and V (bottom) band, respectively. In each panel, the red star points are the original data, the blue triangle points are smoothed data, and the black solid line is the fitting result. The horizontal violet line is a constant background flux, violet dashed lines show the fitted individual flares. The lower panel represents the residual distribution between the fitting data and the original data.}
  \label{fig:distribution}
 \end{figure*}

 Here, we developed the method of automatic fitting of the micro-variability in blazar at the given frequency, and it was achieved by a set of IDL codes compiled in this study based on previous work by \cite{Bha13}. The automatic processing program includes smoothing the light curves, fitting and deducting the base flow, estimating the number of flares, generating the flare parameters by iterative fitting, and generating the turbulent cell parameters and statistical results. The number of flares are estimated from the number of local maximum points and inflection points on the light curves. The flare parameters are calculated on the base of the data points between two local minimum on the light curves and then are used as fitting initial values of the flares. In the program code, the flares are rejected if the data points are less than 10 or the goodness of fit, as represented by correlation coefficient, is less than 99.5\%. 

 In this method, the original data is smoothed automatically before fitting to improve the signal-to-noise ratio of peak information. This smoothing process does not introduce new peaks and does not affect the accuracy of the flare fitting. The automatic fitting code automatically finds the peak in the light curve according to the initial requirement, and automatically generates the initial fitting value of flares. The fitting information includes the width, amplitude of each peak and their central time scales. Then, the code changes the information of each flare step by step by iteration from the initial value until the convolution curve of each flare reaches the desired goodness of fit. The iteration process depends on the $\chi^{2}$ value, the reduction of the fitting curve and the original data points is used as the convergence criterion. The code can fit the flares in the light curves of any object if the profile of flare to be generated for fitting is given. This work fits the observed data of S5 0716+714 from 2011 to 2018, The full versions of Table \ref{tab:datasample} are only available in electronic form. Column 1 in the table represents the observation date, the second column lists the observation band. Column 3 and column 4 represent the number of observational data points and the number of flares obtained by random fitting for the corresponding IDV light curves, respectively.

\begin{table}
	\centering
	\caption{Examples of statistical results catalog}
	\label{tab:datasample}
	\begin{tabular}{cccc}
		\tableline\tableline
		Observation Date & Band & Amount of data & Number of flares \\\tableline
		05-Jan-2016    & I & 241 &  10    \\
		& R & 245 &  11    \\
		& V & 243 &  10    \\\tableline
		06-Jan-2016    & I & 232 &  13    \\
		& R & 234 &  12    \\
		& V & 236 &  10    \\\tableline\tableline
	\end{tabular}
    \tablecomments{\textbf{Table \ref{tab:datasample} is published in its entirety in the machine-readable format. A portion is shown here for guidance regarding its form and content.}}
\end{table}

 It should be noted that the flares obtained by automatic fitting are randomly distributed, so the mid-point of flares obtained in different bands may be different. If the aim is to compare the flare information of the same spatial position in different bands, the initial conditions of the same position can be set in the program. This data processing method will be used in the time delay analysis later.

 An example of model fitting results for S5 0716+714 are shown in Figure \ref{fig:distribution}, which was observed on January 5, 2016. The original light curves and fitting results in I, R and V bands are given. In each panel, the red star points are the original data, the blue triangle points are smoothed data, and the black solid line is the fitting result. The horizontal violet line is a constant background flux, violet dashed lines show the fitted individual flares. The lower panel represents the residual distribution between the fitting data and the original data. The fitting correlation coefficient of this example is higher than 99.73\%.

 Various injection rates, flare times, and flare widths were used to model the IDV light curves in Figure \ref{fig:distribution} are listed in Table \ref{tab:example}. The full versions of Table \ref{tab:example} are published as online table. As mentioned above, the new peaks of cubic spline interpolation are not included in the statistical results. The first and second column of the table lists the index number of the flare and observation band, column 3 and 4 presents the center and the amplitude of the flares. Column 5 gives the half width of the flare, which translates to the time of enhanced injection. The correlation coefficients estimated by the micro-variability are greater than 99.73\%. The amplitude of the flare is then converted to the enhanced injection rate of electrons in the cell as listed in the column 6. Cell sizes are computed by multiplying the time of the flare by shock velocity, and are presented in column 7.

\begin{table}
	\centering
	\caption{Flare parameters used to fit the data}
	\label{tab:example}
	\begin{tabular}{ccp{0.1\columnwidth}p{0.1\columnwidth}p{0.1\columnwidth} p{0.13\columnwidth}p{0.05\columnwidth}}
		\tableline\tableline
		Flare         & $Band$        & $Center$ (hr)         & $Amp$ (mJy)         & $\tau_{flare}$ (hr)         & N $\times$ $10^{-5}$ ($s^{-1} m^{-3}$)         & $S_{cell}$ (AU)         \\\tableline
		1  & I & 23.157 & 0.41 & 0.93 & ~3.17  & 11.08 \\
		& R & 23.035 & 0.29 & 0.94 & ~2.67  & 11.23 \\
		& V & 23.137 & 0.22 & 0.62 & ~2.23  & 7.41  \\\tableline
		2  & I & 24.487 & 1.17 & 0.66 & ~9.02  & 7.87  \\
		& R & 24.427 & 0.87 & 0.73 & ~7.89  & 8.67  \\
		& V & 24.434 & 0.97 & 0.73 & ~9.95  & 8.67  \\\tableline
		3  & I & 25.057 & 1.31 & 0.81 & ~10.01 & 9.67 \\
		& R & 25.005 & 0.33 & 0.49 & ~2.98  & 5.90 \\
		& V & 24.997 & 0.56 & 0.59 & ~5.76  & 7.01 \\\tableline
		4  & I & 25.668 & 0.96 & 1.12 & ~7.40  & 13.27\\
		& R & 25.383 & 1.10 & 1.75 & ~9.96  & 20.81 \\
		& V & 25.494 & 0.74 & 1.62 & ~7.55  & 19.26 \\\tableline
		5  & I & 26.710 & 1.32 & 0.98 & ~10.1  & 11.70 \\
		& R & 26.725 & 1.10 & 1.05 & ~10.00 & 12.53 \\
		& V & 26.746 & 0.77 & 1.00 & ~7.86  & 11.94 \\\tableline
		6  & I & 27.587 & 2.07 & 1.02 & ~15.88 & 12.08 \\
		& R & 27.545 & 1.74 & 0.85 & ~15.71 & 10.07 \\
		& V & 27.440 & 1.14 & 0.74 & ~11.68 & 8.76 \\\tableline
		7  & R & 28.058 & 1.37 & 0.62 & ~12.37 & 7.39 \\
		& V & 28.031 & 1.38 & 0.95 & ~14.06 & 11.35 \\\tableline
		8  & I & 28.309 & 1.43 & 0.71 & ~11.01 & 8.50 \\
		& R & 28.445 & 0.94 & 0.56 & ~8.50  & 6.72 \\
		& V & 28.507 & 0.50 & 0.45 & ~5.15  & 5.42 \\\tableline
		9  & I & 29.428 & 2.61 & 2.16 & ~20.08 & 25.68\\
		& R & 29.469 & 2.27 & 1.96 & ~20.49 & 23.26 \\
		& V & 29.498 & 1.96 & 1.89 & ~20.01 & 22.39 \\\tableline
		10 & I & 31.001 & 2.10 & 1.25 & ~16.15 & 14.82\\
		& R & 31.007 & 1.88 & 1.33 & ~17.01 & 15.78 \\
		& V & 30.911 & 1.63 & 1.08 & ~16.66 & 12.81 \\\tableline
		11 & I & 31.822 & 1.52 & 0.84 & ~11.67 & 10.04 \\
		& R & 31.724 & 0.76 & 0.56 & ~6.87  & 6.70 \\\tableline
		\hline
	\end{tabular}
    \tablecomments{\textbf{Table \ref{tab:example} is published in its entirety in the machine-readable format. A portion is shown here for guidance regarding its form and content.}}
\end{table}

\section{Result and Discussion}

 We developed a method of fitting the micro-variability model describes in Sect. 3.1 and 3.2, that automatically fits a light curve, and results in the relevant parameter e.g injection parameter. With the automatic fitting, we can quickly obtain the amplitude, duration and central position of all the peaks. If there are $N$ flares (the number of flares is automatically calculated by code), the parameters of $(3N+1)$ degrees of freedom can be automatically obtained. It should be noted that $(9N+1)$ degrees of freedom parameters can be obtained for the light curves of three bands in the same period if the flares at the same time among different bands are assumed to be generated by the same turbulence. Statistical analysis of the data obtained from the method are described below.

\subsection{Fitting result and discussion}

 We use 75 simultaneous light curves of the object S5 0716+714. Finally, 332 flares with reliable goodness of fit and similar distribution among different bands were selected, including 110 in I band, 115 in R band and 107 in V band. The fitting results are analyzed as follows.

\begin{figure*}
   \begin{minipage}{\textwidth}
   \centering
   \includegraphics[trim=0.6cm 0.5cm 0cm 1.8cm,width=0.45\textwidth,clip]{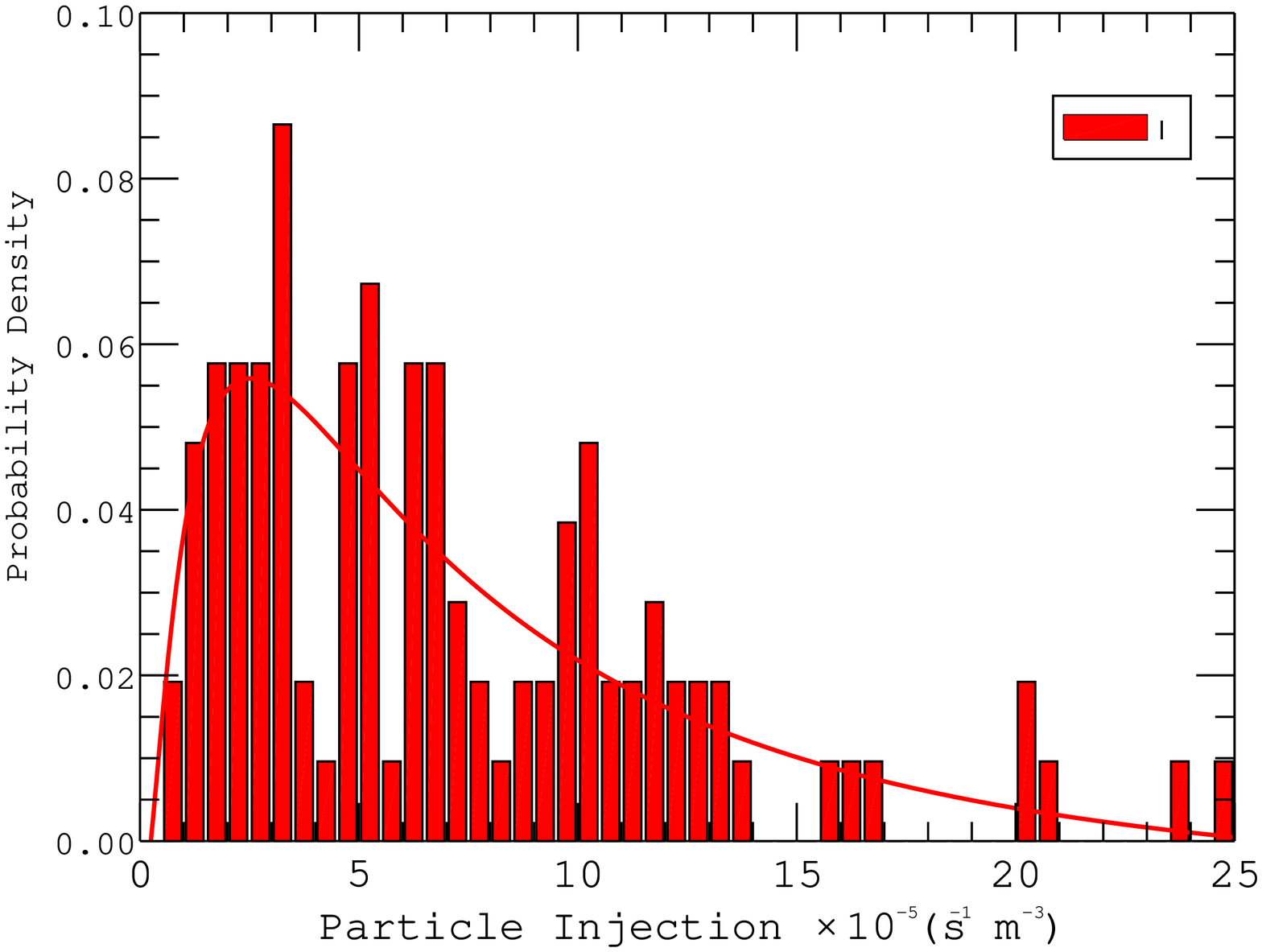}
   \includegraphics[trim=0.6cm 0.5cm 0cm 1.8cm,width=0.45\textwidth,clip]{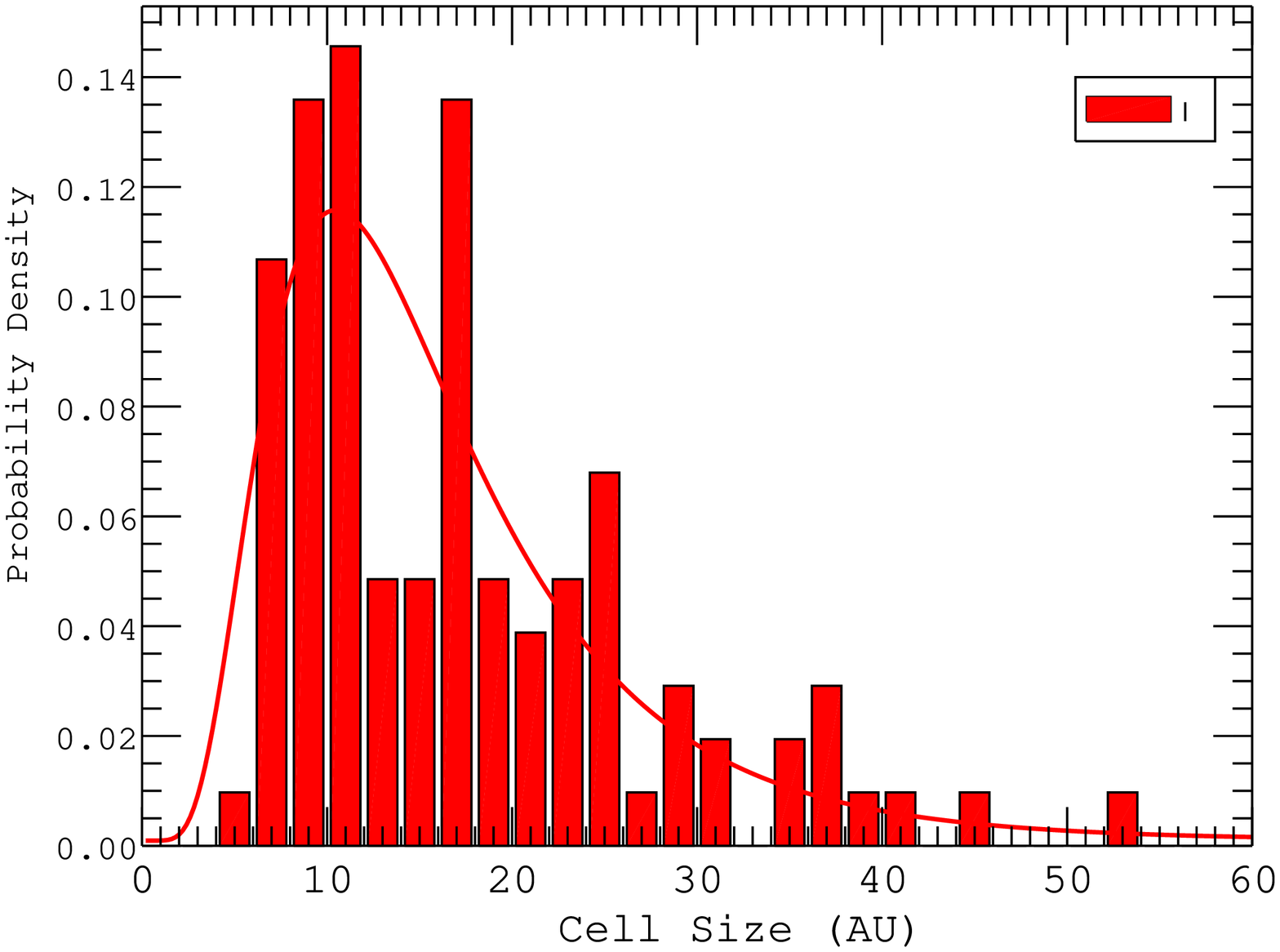}
   \includegraphics[trim=0.6cm 0.5cm 0cm 1.8cm,width=0.45\textwidth,clip]{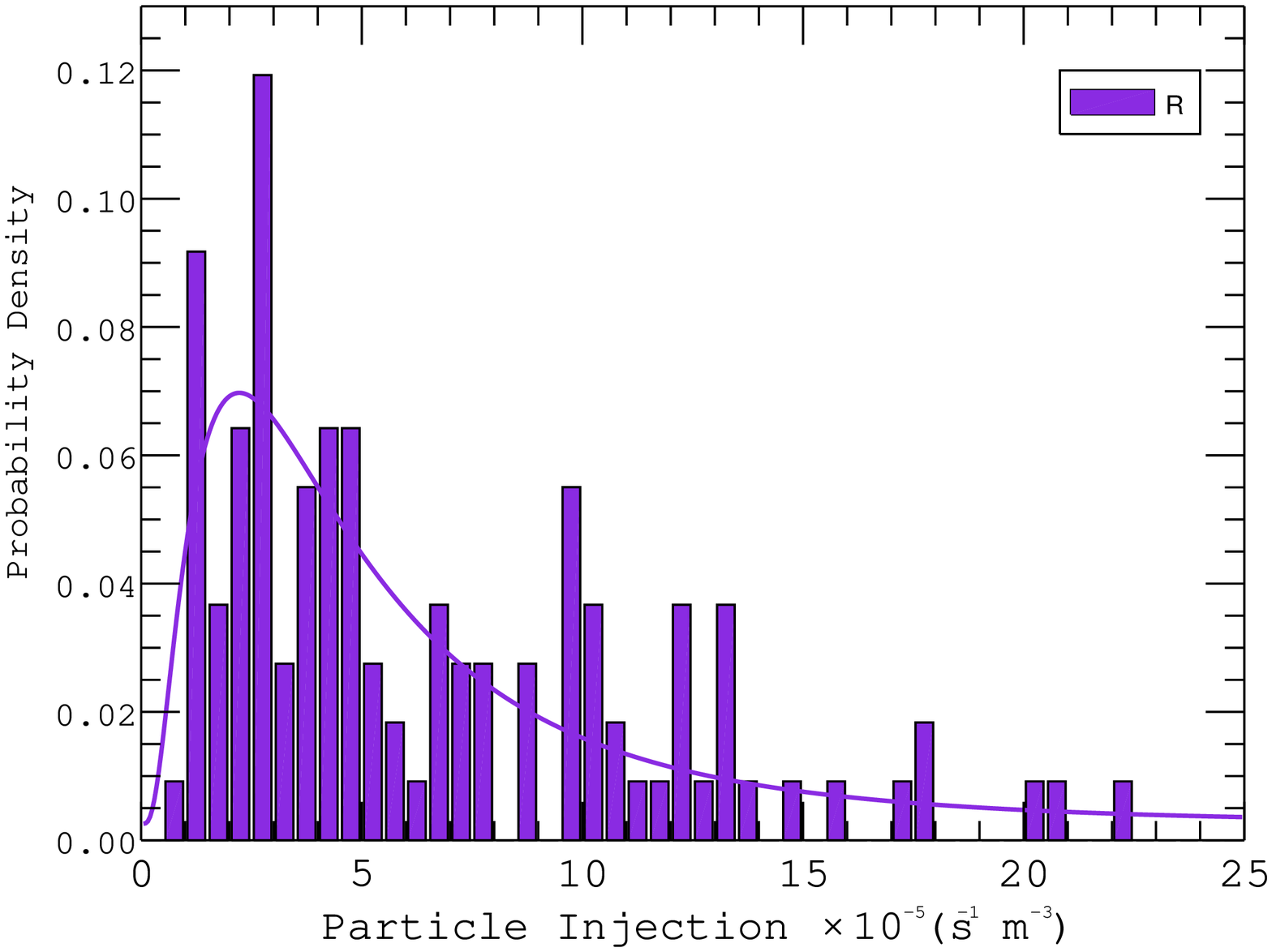}
   \includegraphics[trim=0.6cm 0.5cm 0cm 1.8cm,width=0.45\textwidth,clip]{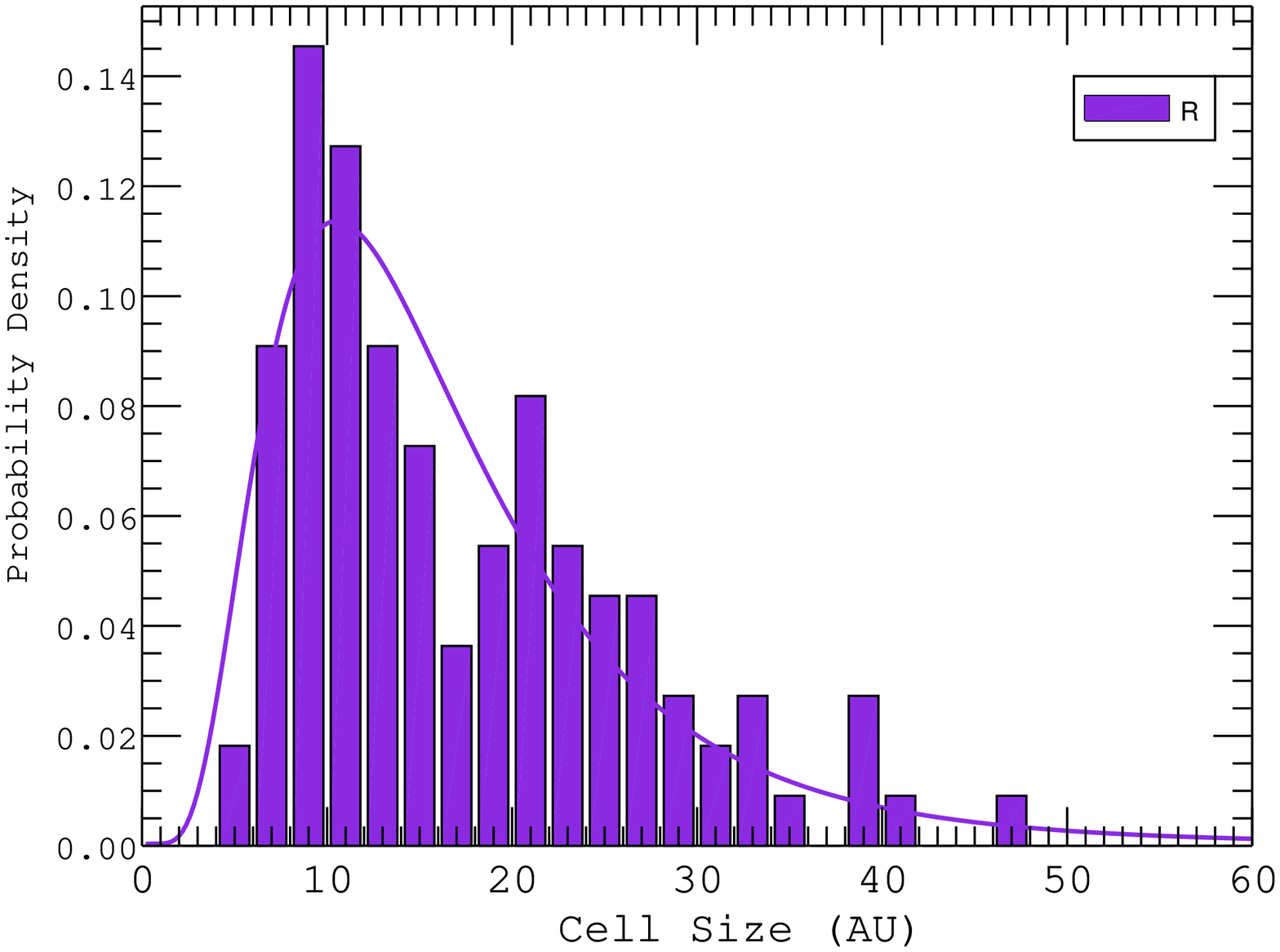}
   \includegraphics[trim=0.6cm 0.5cm 0cm 1.8cm,width=0.45\textwidth,clip]{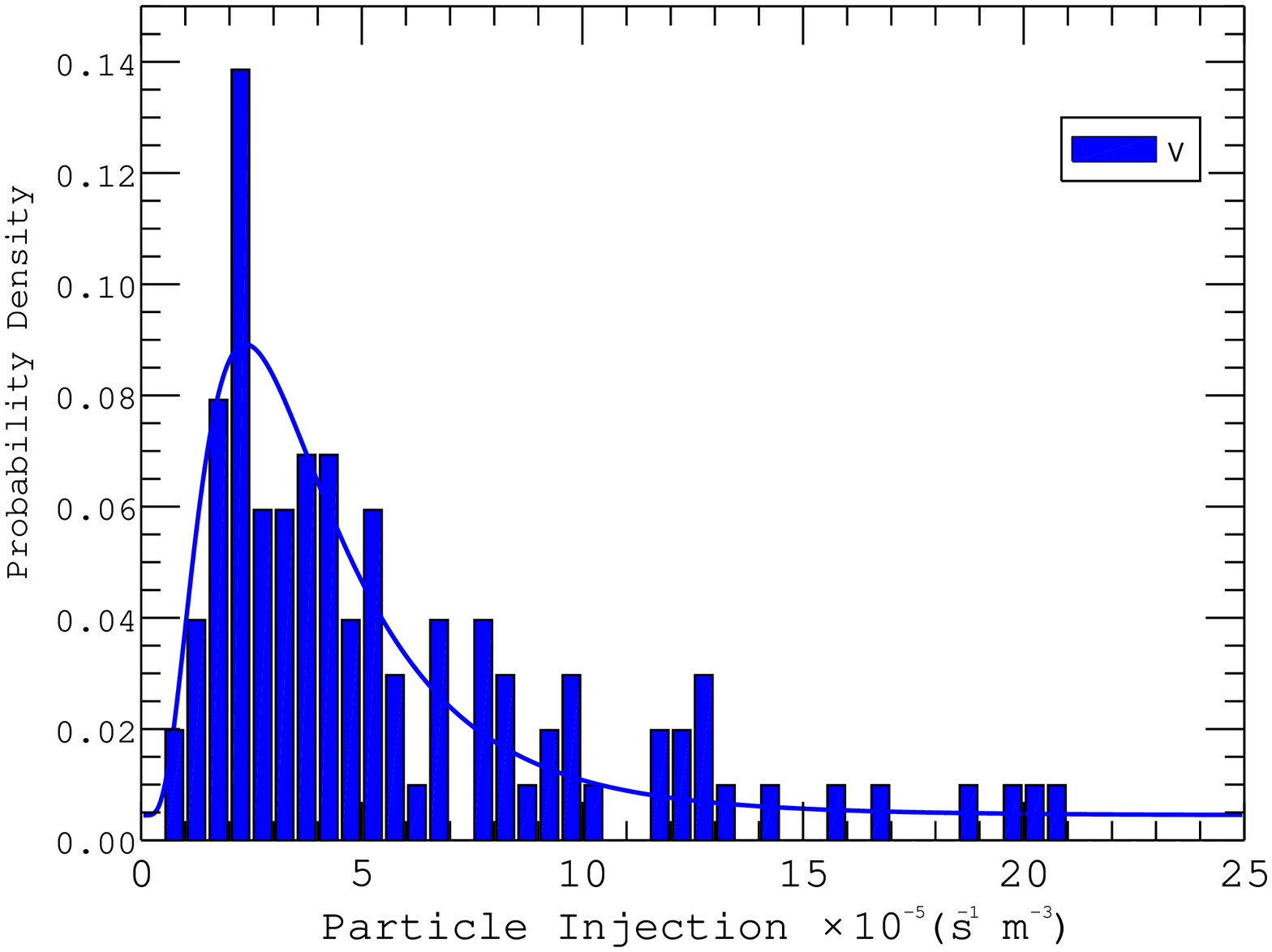}
   \includegraphics[trim=0.6cm 0.5cm 0cm 1.8cm,width=0.45\textwidth,clip]{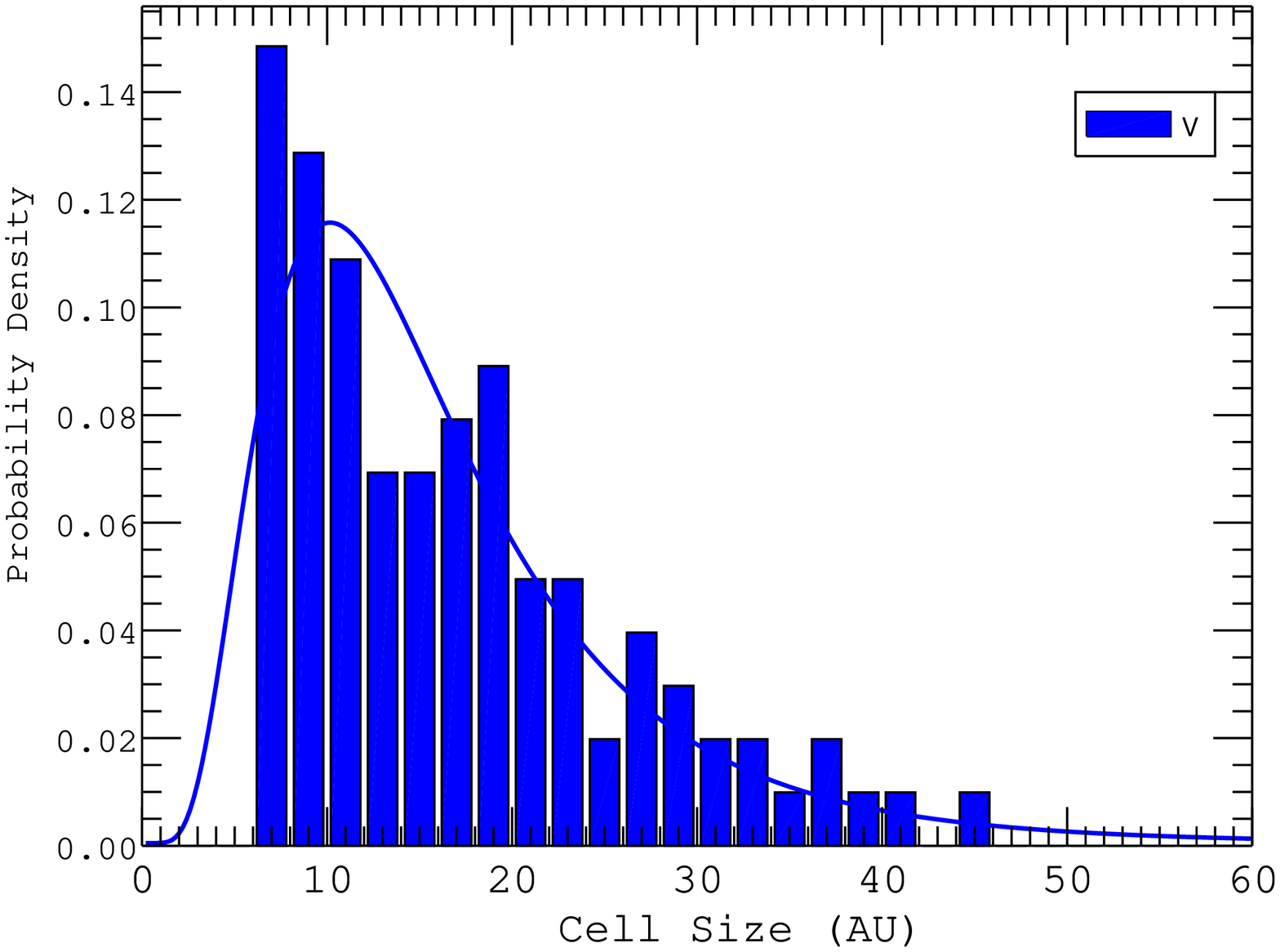}
   \end{minipage}\vspace{0.001cm}
   \caption{Statistical and distribution diagrams of cell size and particle injection rate. Red, violet and blue solid lines represent the log-normal distribution corresponding to the I, R and V bands, and histograms show the probability density of cell size and particle injection.}
   \label{fig:lognom}
 \end{figure*}

 To investigate the distribution of amplitude and duration time, a two-parameter probability density function of the log-normal distribution is represented as follows \citep{Ait57,Cro88}
 \begin{equation}
   p(x) = \frac{1}{x\sigma \sqrt{2\pi}}e^{-\frac{(ln x-\mu)^{2}}{2\sigma^{2}}}
 \end{equation}
 The log-normal distribution fitting of turbulent cell size and particle injection rate is given in Figure \ref{fig:lognom}. Red, violet and blue solid lines represent the log-normal distribution corresponding to the I, R and V bands, and histograms show the probability density of cell size and injection rate. According to the distribution characteristics, almost all turbulent cell sizes are between 5 and 55AU, and the value of particle injection rate is not zero. It shows that the micro-variability of this object has a very large duty cycle.

 Since the generation of turbulent cell usually is a stochastic process, every micro-variability light curve is the realization of the stochastic process. There is a wide range of the turbulent cell sizes. The smallest scale of the turbulent cell usually is related to the Kolmogorov scales, and most of the dissipation occurs in non-relativistic plasma. Viscosity dominates and the turbulent kinetic energy is dissipated into heat at the Kolmogorov scale. The Kolmogorov length scale and velocity scale are defined by
 \begin{equation}
   \eta = \left(\frac{\nu^{3}}{\varepsilon} \right )^{1/4}
 \label{equ:Kollength}
 \end{equation}
 \begin{equation}
   u_{\eta} = \left( \nu \varepsilon \right )^{1/4}
 \label{equ:Kolvel}
 \end{equation}
 where $\varepsilon$ is the average rate of dissipation of turbulence kinetic energy per unit mass, and $\nu$ is the kinematic viscosity of the fluid. The Kolmogorov length scale can be obtained as the scale at which the Reynolds number is equal to 1.
 \begin{equation}
   Re = \frac{\eta u_{\eta}}{\nu} = 1
 \label{equ:Reynolds}
 \end{equation}
 where $Re$ is Reynolds number. Based on the smallest size obtained by model fitting, the viscosity of the plasma material in the jet \citep{Smi12} can be estimated by using the velocity scale of turbulence in relativistic jets, and vice versa. In this model fitting, the smallest scale \citep{Fal06} is calculated as $\sim 5 AU$. It is considered that the minimum turbulence scale corresponds to the critical Reynolds value. If the Reynolds value is higher than the critical value, that means the plasma flow is turbulent than laminar on the short-term scale. Figure \ref{fig:lognom} shows that the size of largest cell is within $\sim 55 AU$. We suppose that the maximum scale is related to the size of plasma jet or correlation length, and turbulent cells beyond this scale may become unstable. Based on the fact that the micro-variability time scale should be less than one day, so the turbulent structure with light-travel time greater than one day should not be included in this work.

 \begin{figure*}
	\begin{minipage}{\textwidth}
		\centering
		\includegraphics[trim=0.6cm 0.8cm 1.7cm 1.8cm,width=0.42\textwidth,clip]{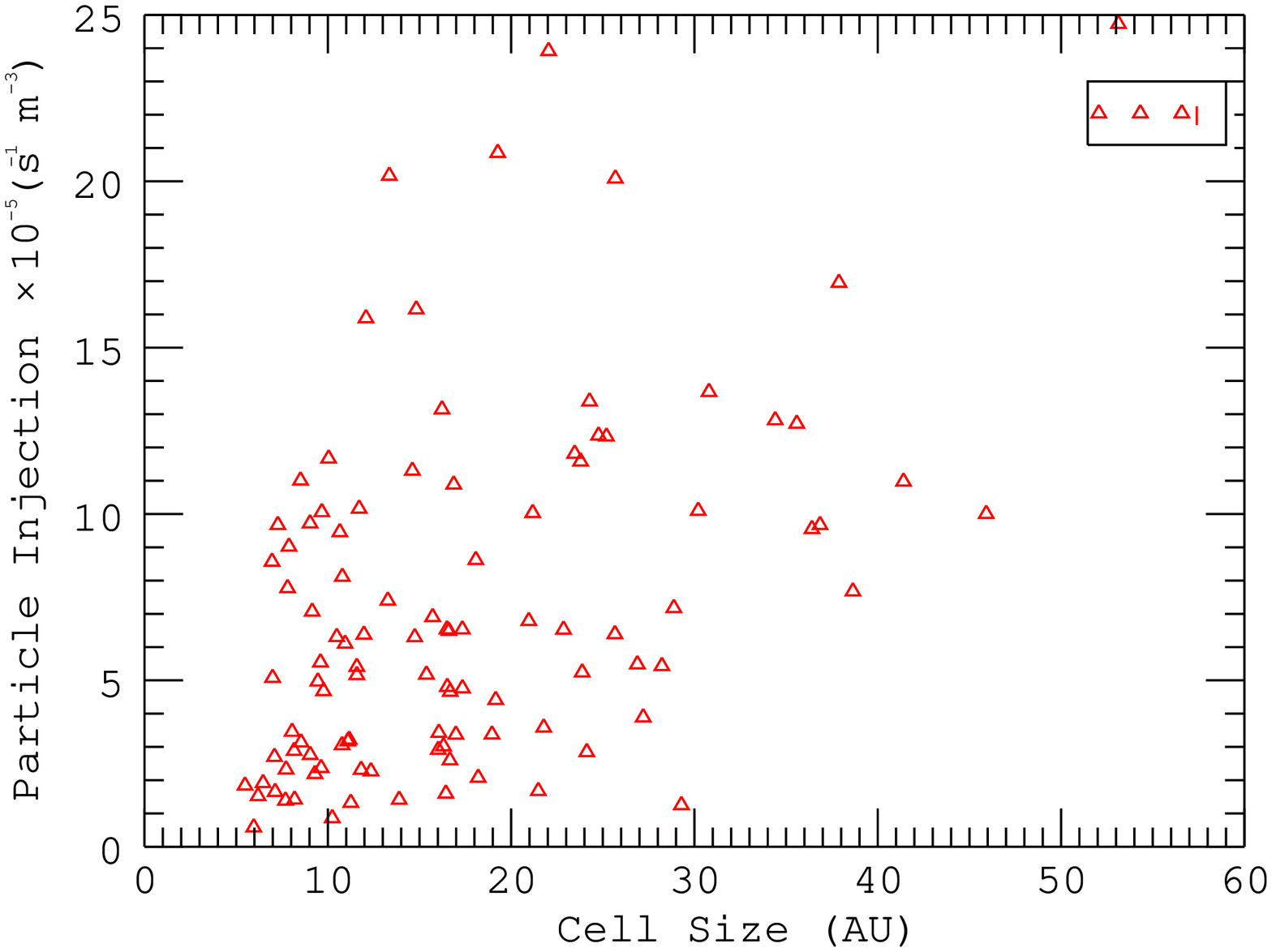}
		\includegraphics[trim=0.6cm 0.8cm 1.7cm 1.8cm,width=0.42\textwidth,clip]{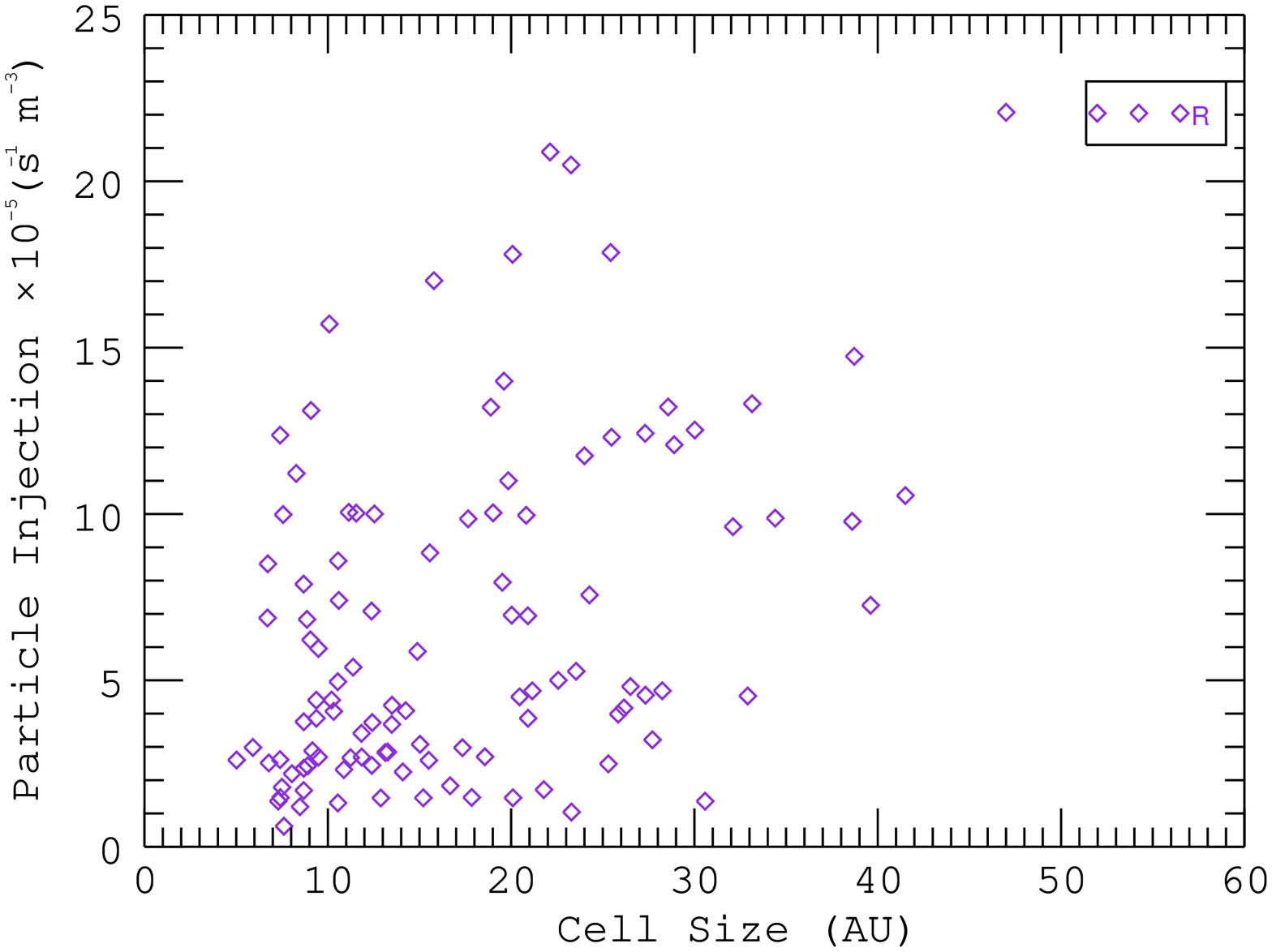}
		\includegraphics[trim=0.6cm 0.8cm 1.7cm 1.8cm,width=0.42\textwidth,clip]{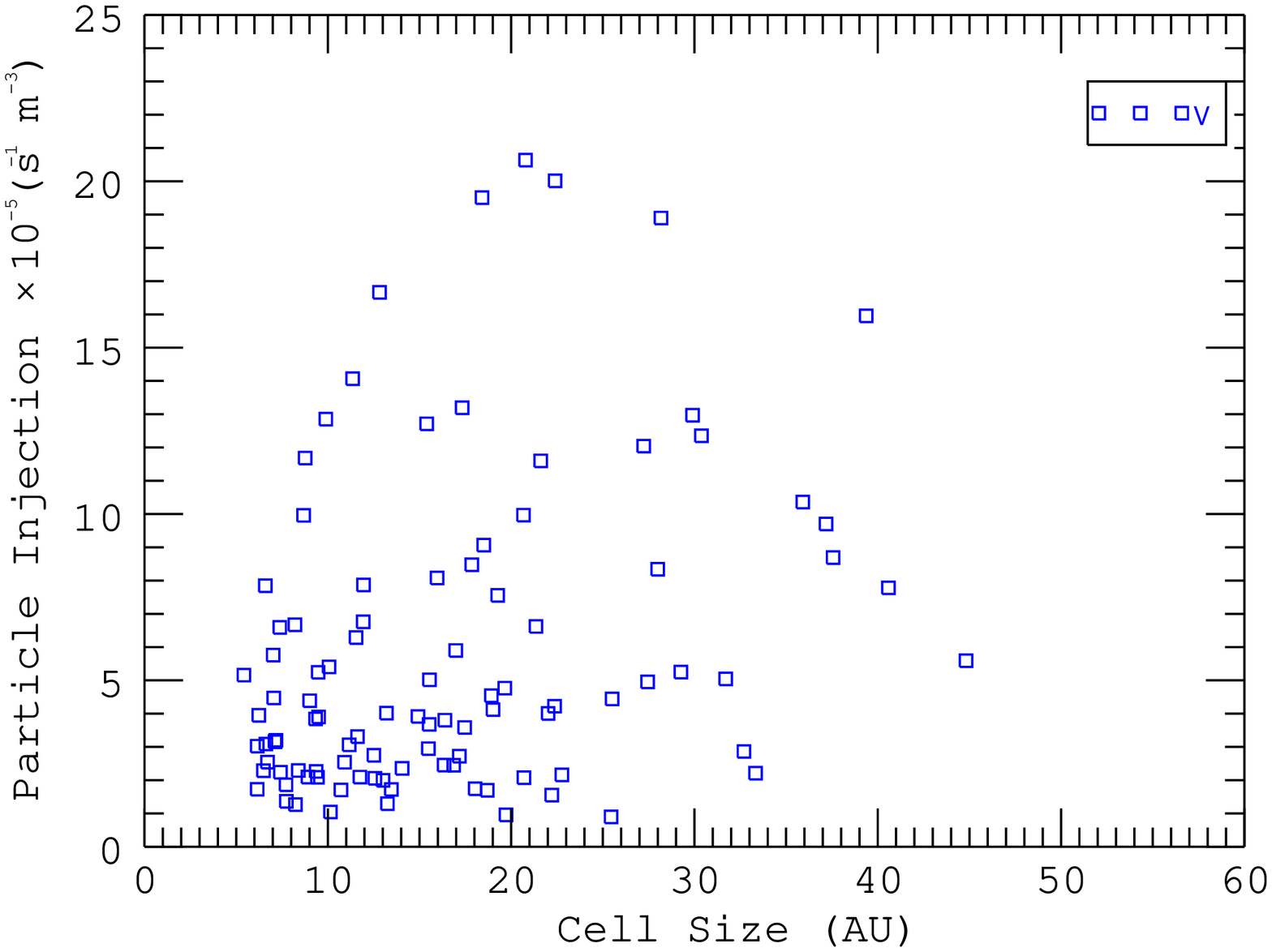}
	\end{minipage}\vspace{0.001cm}
	\caption{The distribution of particle injection rates in the cells versus turbulent region cell sizes. The triangle point represents I band, the rhombus point represents R band, and the square point represents V band.}
	\label{fig:statistical}
\end{figure*}

 In the Figure \ref{fig:statistical}, the comparative distributions of cell particle injection rate and turbulent cell size are given. This picture shows that there is no obvious correlation between them. But they may have similar distribution and range, which needs further research and verification.

\subsection{Time delay analysis}

 It is assumed that flares in I, R and V bands at similar times originate from the same turbulent cell. According to the KRM equation, the sizes of turbulent cells and the density of particles at different frequencies in the optical band should be slightly different. Based on this assumption, the duration should be almost the same in different bands, and the particle injection rate obtained by fitting should be slightly different. By setting the initial conditions in the fitting procedure, the flare parameters with similar central time scales are obtained. The difference between the central time of the flares at the same position in different bands is the time lag. In this statistic, the initial fitting condition discards the flare whose central time difference is greater than 1 hour at the same position at different frequencies. After adding this screening condition, the remaining number of flares is still more than 95.48\% of the total statistics. Time lag analysis is obtained from these remaining high-quality fitting results.

 \begin{figure*}
	\begin{minipage}{\textwidth}
		\centering
		\includegraphics[trim=0.1cm 0.6cm 1.7cm 1.8cm,width=0.42\textwidth,clip]{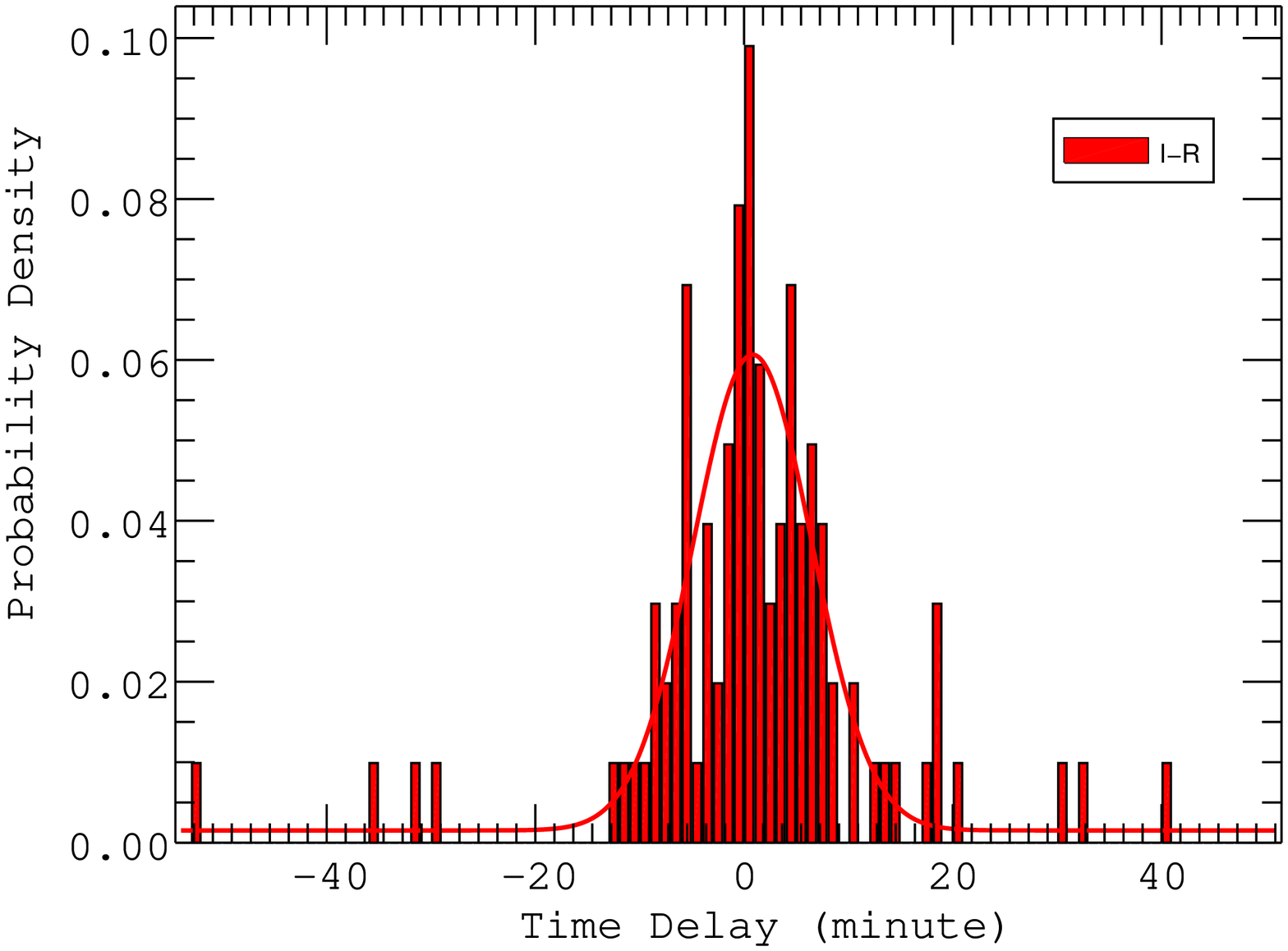}
		\includegraphics[trim=0.1cm 0.6cm 1.7cm 1.8cm,width=0.42\textwidth,clip]{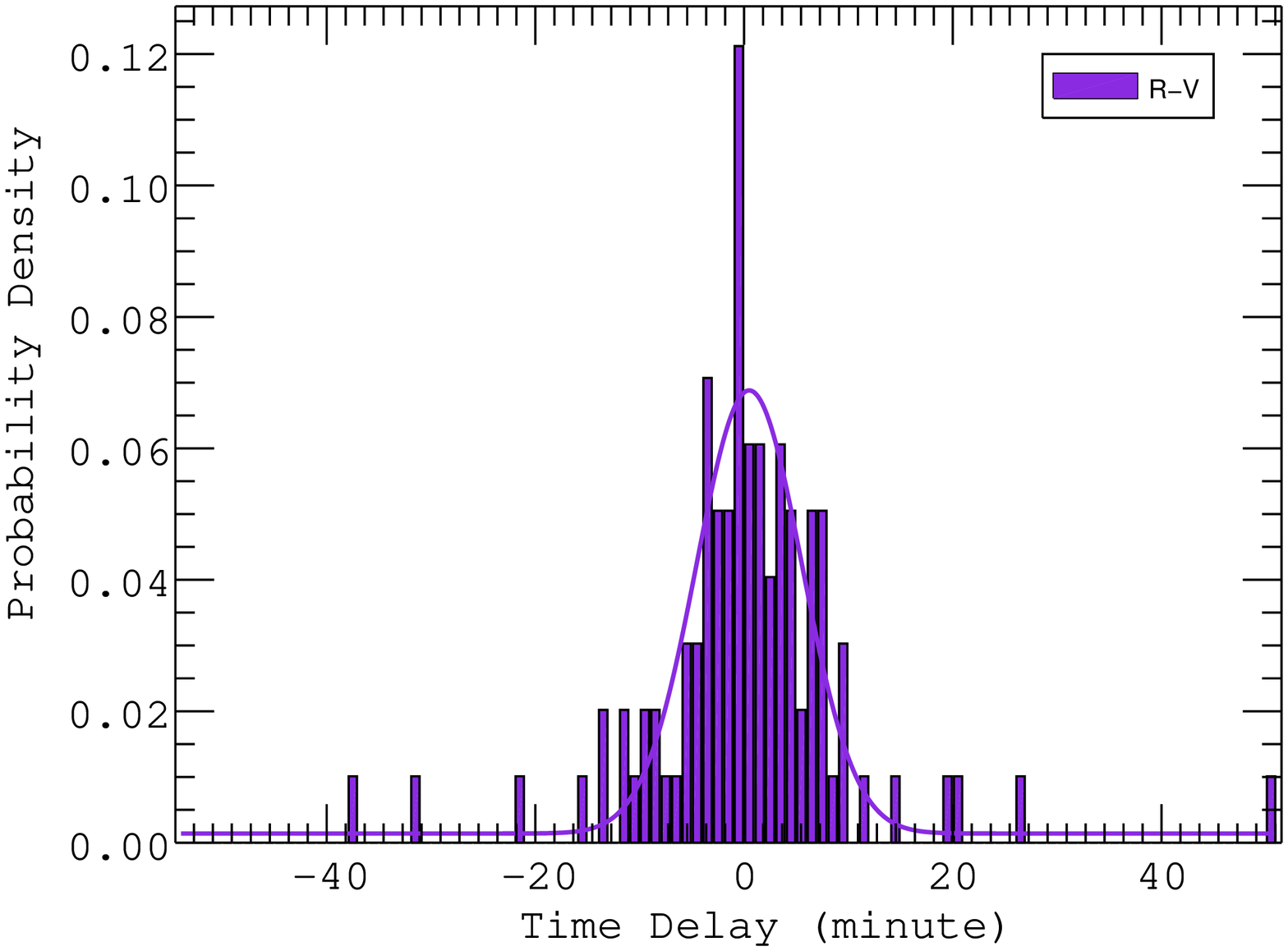}
		\includegraphics[trim=0.1cm 0.6cm 1.7cm 1.8cm,width=0.42\textwidth,clip]{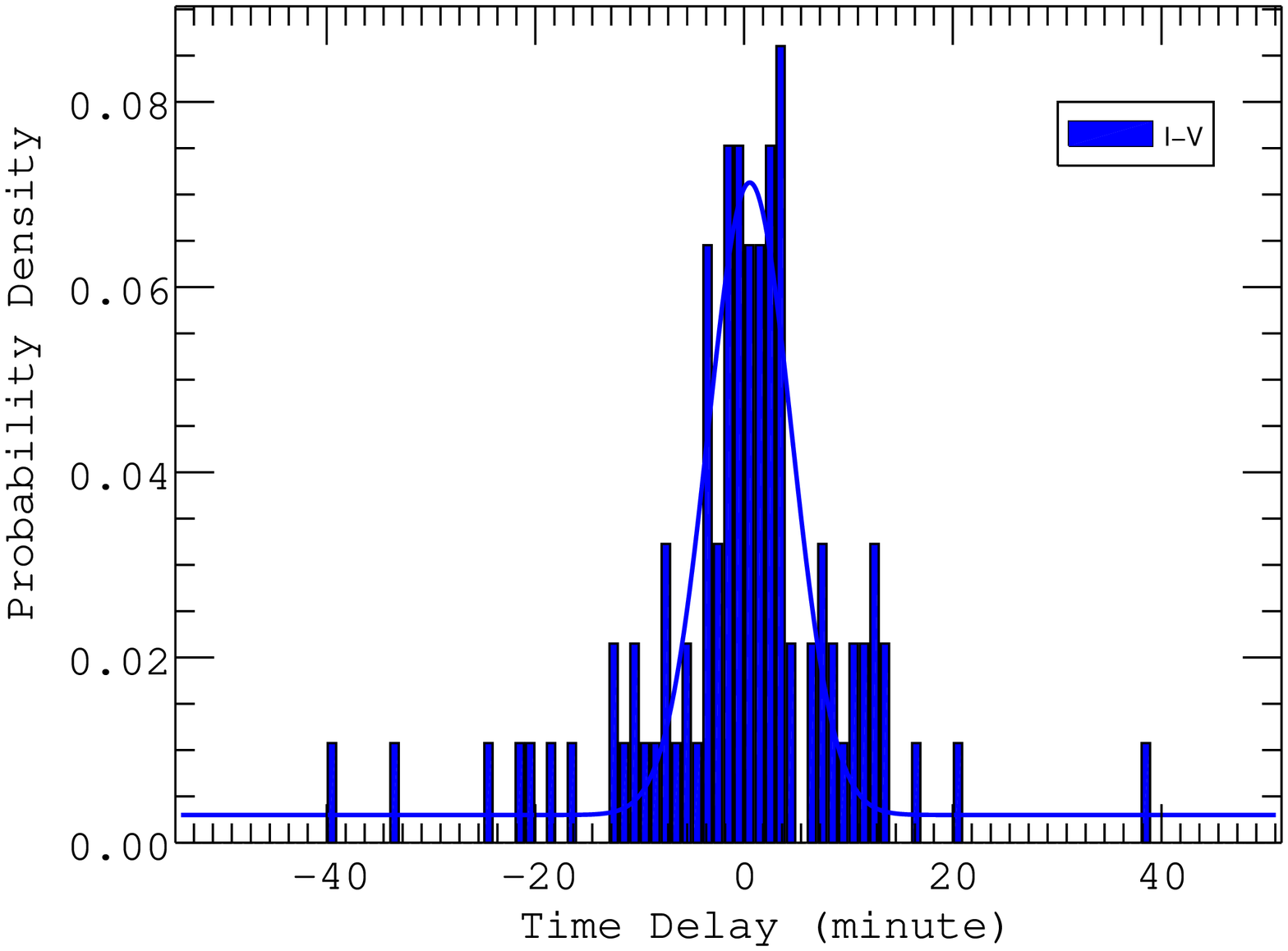}
	\end{minipage}\vspace{0.001cm}
	\caption{The distributions of the time lag for micro-variability light curves are between I and R, R and V, V and I bands, respectively. The red, violet and blue solid lines represent the normal distribution fitting of time delay between I and R, R and V, V and I bands, respectively.}
	\label{fig:nom}
\end{figure*}

 According to the theoretical analysis of the model, two flares with similar central time scales at different frequencies should be generated by the same turbulent cell, so the difference of the central time scale is used to express the time lags between different frequencies. By analyzing the correlation of the whole light curve, we prove that the correlation of the similar flares is the highest when they are generated in the same cell. In Figure \ref{fig:nom}, the histogram distribution of the time delay between similar flares of I and R, R and V, V and I bands is presented. The red, violet and blue solid lines represent the normal distribution fitting of time delay between I and R, R and V, I and V bands, respectively. Histogram shows the probability distribution of time delay at different frequencies. The statistical results show that the time lags between the micro-variability flares in different optical bands are about several minutes. This is also consistent with the previous assumption that flares at different frequencies turbulent generated at the same location. According to the statistical results, the time delay between I and R is about $0.81 \pm 11.34$ minutes, R and V is about $0.49 \pm 10.11$ minutes, I and V is about $0.54 \pm 8.51$ minutes, respectively. We obtained the log-normal distribution of flare duration time scales, and it can explain why the time delays also conform to a normal distribution. The reason why flare duration time scales and time delays conform to a normal distribution is unclear, so we need further investigations of generation and distribution of turbulence.

 For the KRM equation, the flare duration scale is correlated with time delay, but it is not strictly positive correlation. Based on this assumption, the time dependence of flare convolution may be worse due to the difference of turbulent cell size. This phenomenon may imply that the time delay of the whole light curve is the weighted average of the time delay of each flare. In order to compare the relationship and difference between global and local time delays, we need to solve the time delays of the whole light curves.

 The cross-correlation function (CCF) is the standard tool used to measure the time lag. We use the local cross-correlation function (LCCF) to calculate the correlation and time lags of the light curves at different frequencies \citep{Wel99}, and then compare the relationship between the overall time delay and the time delay of each flare. The standard definition of the LCCF of two time series $x_{i}$ and $y_{i}$ sampled at only those (N - k) points that overlap at a given lag ($\tau_{k} = k\triangle t$) are used to determine the mean and standard deviations is
 
 \begin{equation}
 \begin{split}
   LCCF(\tau_{k})\equiv  & \frac{[1/(N-k)]\Sigma_{i=1}^{N-k}(x_{i}-\bar{x_{*}})(y_{i+k}-\bar{x_{*}})} {\left \{ \left [1/(N-k)\Sigma_{i=1}^{N-k}(x_{i}-\bar{x_{*}})^{2} \right]^{1/2} \right \}} \\
   & \times \left \{ \left [1/(N-k)\Sigma_{i=k+1}^{N}(y_{i}-\bar{y_{*}})^{2} \right ]^{1/2} \right \}
 \end{split}
 \end{equation}
 
 where
 
 \begin{equation}
   \bar{x_{*}} = \frac{1}{N-k} \Sigma_{i=1}^{N-k} x_{i} ~,~ \bar{y_{*}} = \frac{1}{N-k} \Sigma_{i=k+1}^{N} y_{i}
 \end{equation}
 
 are the means of $x_{i}$ and $y_{i}$ in the overlap interval. The result shows that I, R and V bands are well correlated with each other and the delay time between I and R band is so small (about $0 \sim 14.4min$), sometimes equal to 0 minutes. We compare the time delay obtained by this work with previous studies \citep{Wu12, Bha16}, and find that these time delay obtained are about several minutes. However, the results of LCCF analysis are consistent with those of KRM equation analysis. It shows that the time delay of the whole light curve analysis or the single flare analysis is consistent. The above analysis shows that the time delay between the flares may be one of the reasons for the time delay of the micro-variability light curves in different bands. This hypothesis needs further research to confirm.

 \begin{figure*}
   \begin{minipage}{\textwidth}
   \centering
   \includegraphics[trim=0.4cm 0.5cm 0cm 1.8cm,width=0.45\textwidth,clip]{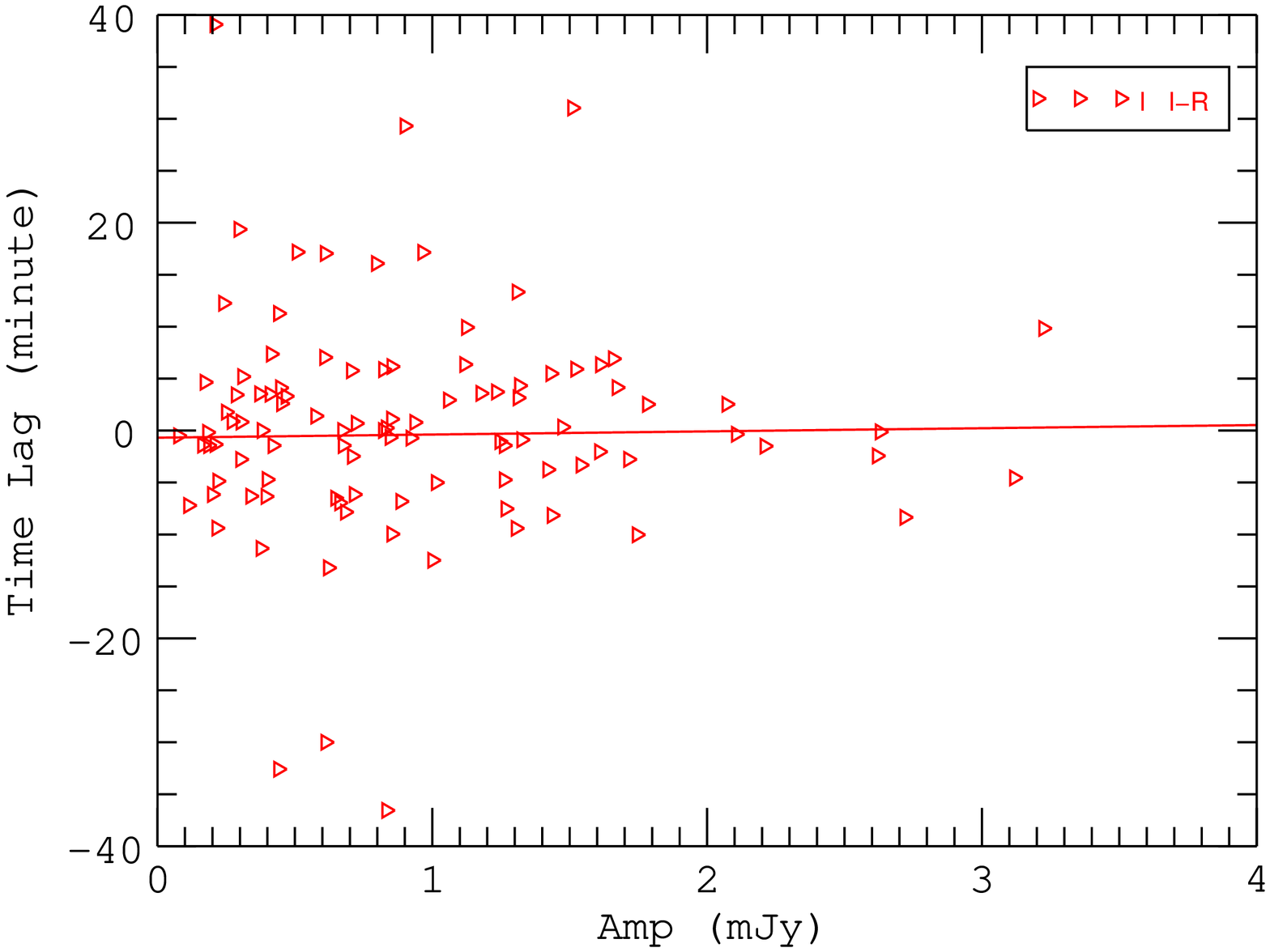}
   \includegraphics[trim=0.4cm 0.5cm 0cm 1.8cm,width=0.45\textwidth,clip]{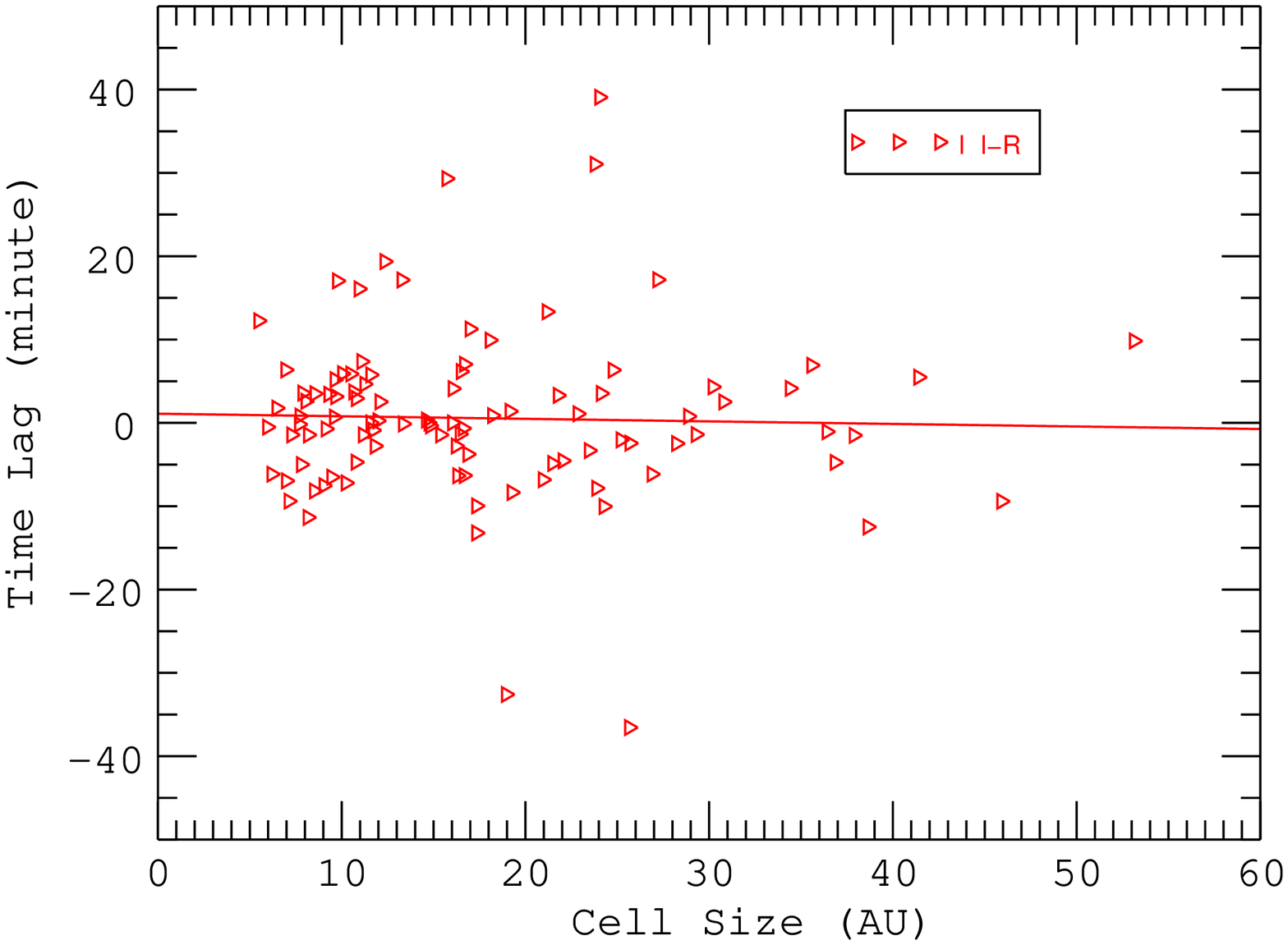}
   \includegraphics[trim=0.4cm 0.5cm 0cm 1.8cm,width=0.45\textwidth,clip]{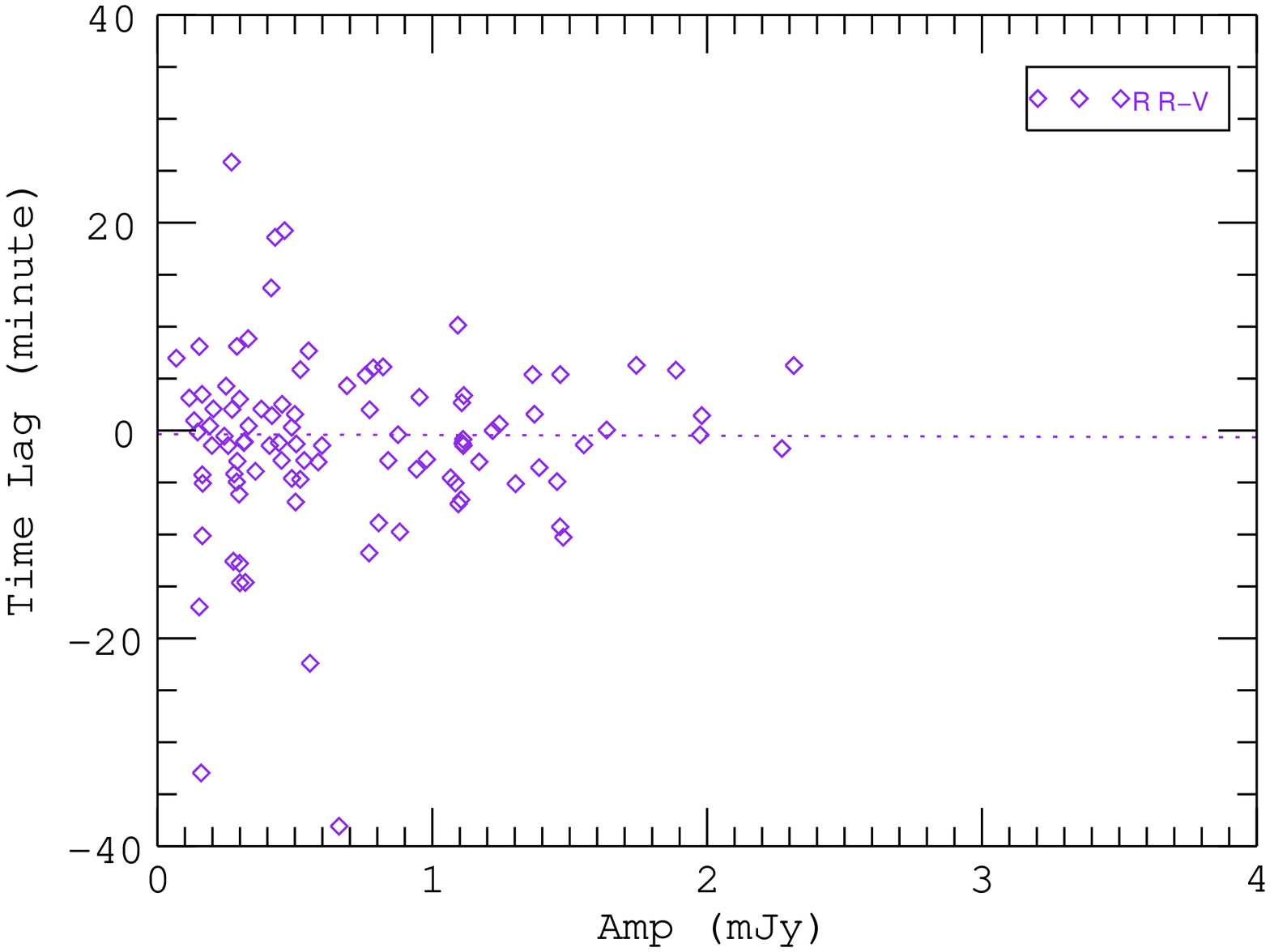}
   \includegraphics[trim=0.4cm 0.5cm 0cm 1.8cm,width=0.45\textwidth,clip]{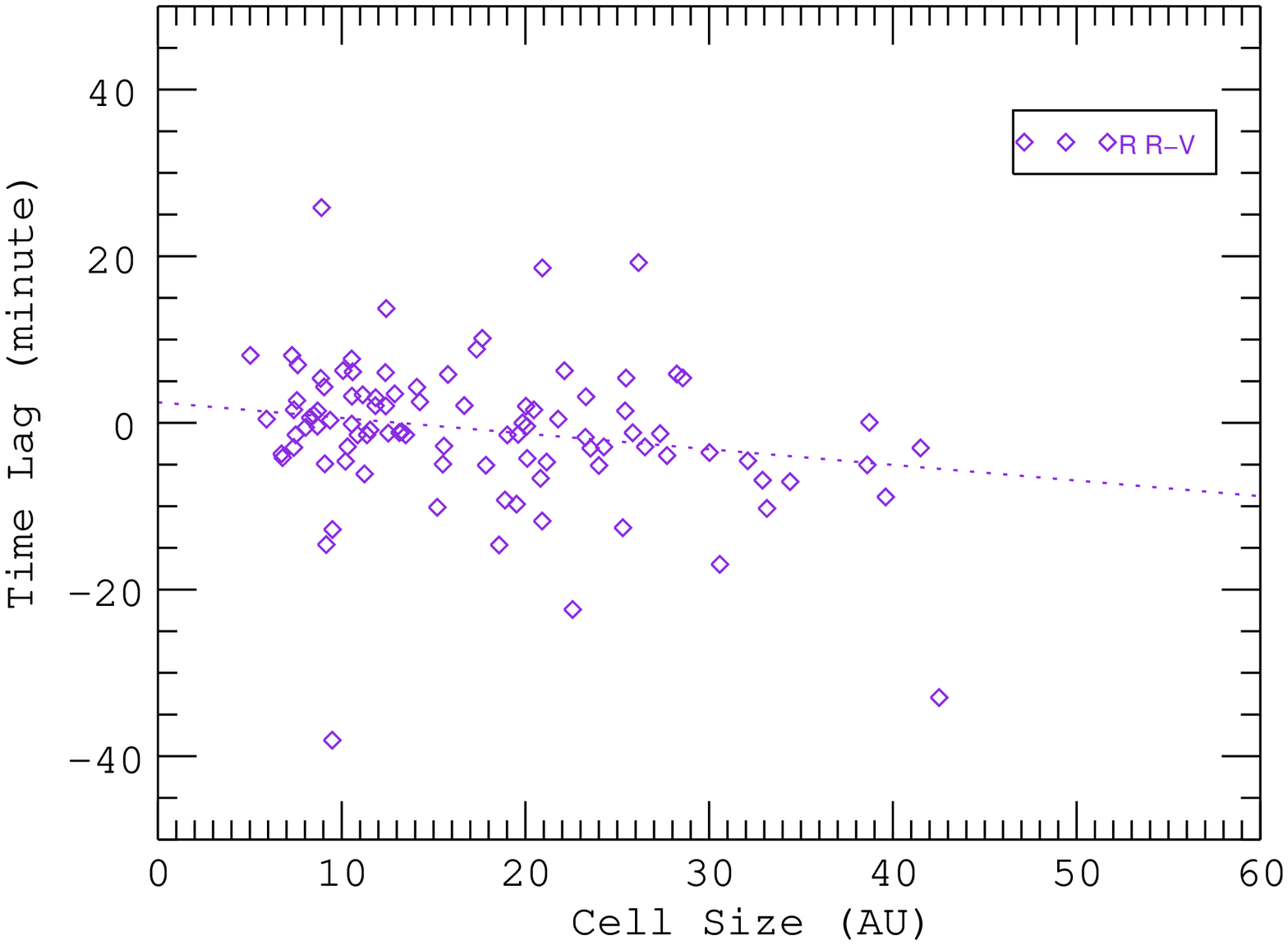}
   \includegraphics[trim=0.4cm 0.5cm 0cm 1.8cm,width=0.45\textwidth,clip]{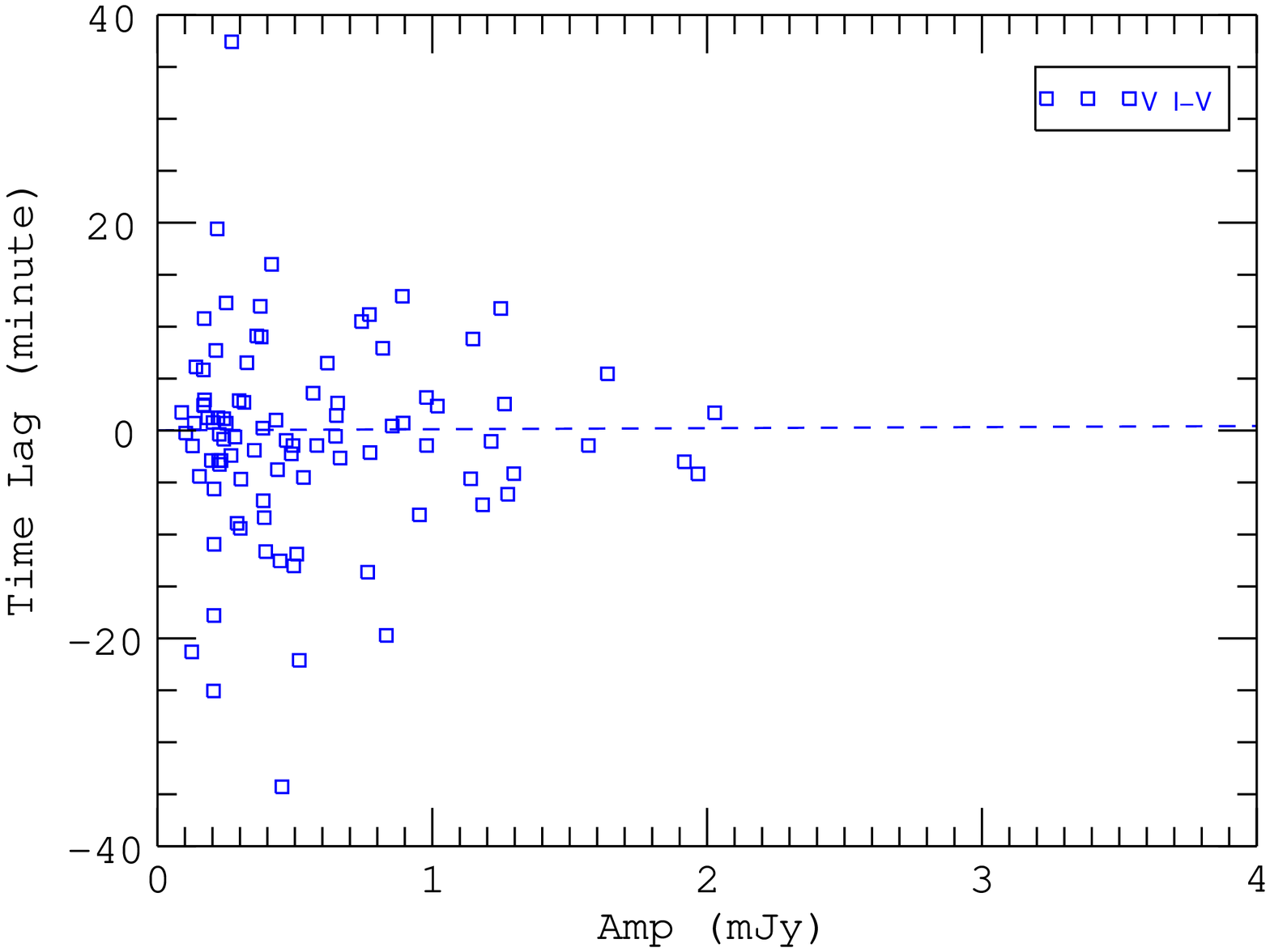}
   \includegraphics[trim=0.4cm 0.5cm 0cm 1.8cm,width=0.45\textwidth,clip]{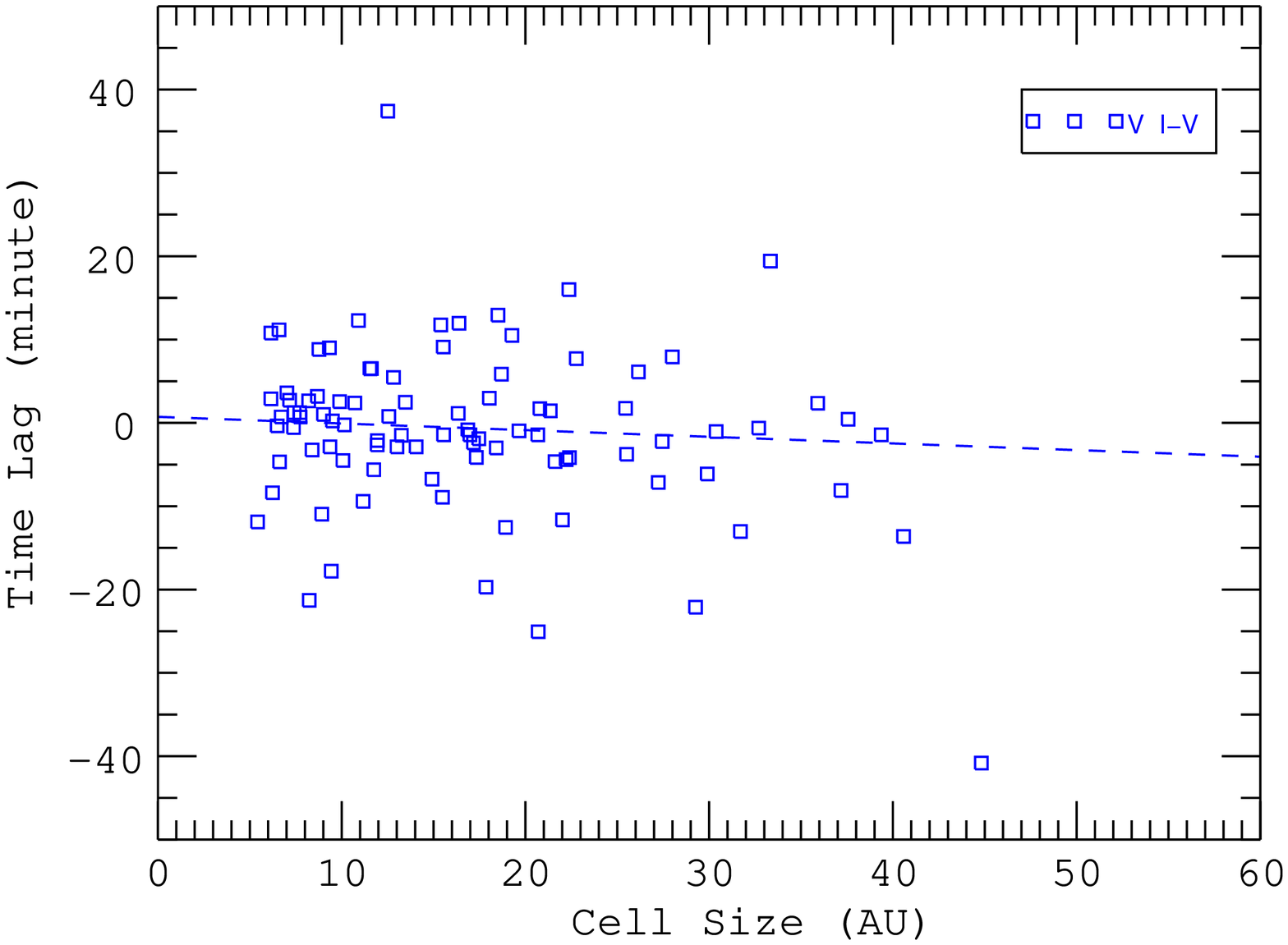}
   \end{minipage}\vspace{0.001cm}
   \caption{Distribution of flux (left) and cell size (right) with time lag, respectively. The triangle, diamond and square points represent the time lags between different frequencies fitted respectively. Solid lines, dashed lines and dotted lines represent the trend of average fitting data, respectively. }
   \label{fig:timedelayfit}
 \end{figure*}

 The relationship between flare amplitude and cell size and time lag is given in Figure \ref{fig:timedelayfit}. The cell size and flare flux are obtained by fitting the micro-variability light curves, and the time lag is based on the difference of the central time scale of each flare at different frequencies. Figure 11 (left panel) shows that at any band the time leg of flares does not depend on amplitude of flares. However, the right panel in Figure \ref{fig:timedelayfit} shows that the time lag of flare increases with the increase of turbulent cell size, which is consistent with the theoretical analysis of KRM equation. It should be noted that both theoretical calculations and simulation results show that the relationship between time lag and flare time scale is not linear. Quantitative analysis of factors affecting time lag needs more observations and data analysis.

\section{Summary and Conclusion}

 Well-sampled micro-variability light curves of S5 0716$+$714 observed by Weihai observatory of Shandong University were analyzed. The model of synchrotron radiation cell in jet proposed by \cite{Bha13} is used. Based on this model and the observation data, the physical mechanism of flare formation in micro-variability and its influencing parameters were systematically analyzed. We developed the method based on KRM equation to automatically fit the micro-variability light curve. The data of blazar S5 0716$+$714, obtained during 8 years, are fitted and counted through the compiled code. We decomposed the observed micro-variability light curve into individual flares, and get the number of flares, central time, duration time and the amplitude of each flare by fitting. By correlating the fitted data with the physical model, the size of turbulent cells, the enhanced particle density and the time lags between different frequencies are calculated.

 According to the statistical results of automatic fitting, the turbulent cell size is within the range of about 5 to 55 AU. It shows that turbulence exists in most regions of the jet, and the Kolmogorov scale can be estimated according to the minimum statistical scale. The statistical results of the enhanced particle injection and scale of turbulent cell are approximately log-normal distribution. However, the time lags between different frequencies show a normal distribution, and 95\% of them are within $\pm 10$ minutes. We also deduce theoretically that the time lag of each flare may be related to its duration, and it is about $12.98_{-12.98}^{+9.23}$ minutes. This comparison shows that the time lag between different frequencies of flares in optical band may be due to the response time difference of the interaction between shock and plasma. The time delay of IDV light curves were calculated by LCCF method, and their average value was about $14.40$ minutes. We found that the time delay of the light curves are similar to that of flares in the certain range. This result may imply that the time lag between the light curves is the weighted average of time lag for each flare.

 In the optical regime, time delay is difficult to detect directly because it depends on high time resolution and measurement accuracy. When the time lags between light curves are too small to be calculated, it is a feasible method to estimate the time lags between light curves by calculating the time lags of flares. For S5 0716$+$714, many investigators have analyzed the time delays \citep{Qia00, Vil00, Poo09, Zha10}. By comparing these results, we found that the time delays between optical bands are about several minutes. The comparison results show that it is possible to estimate the time delay by flares.

\section*{Acknowledgements}
 This work is supported by Natural Science Foundation of China under grant No. 11873035, the Natural Science Foundation of Shandong province (No. JQ201702) and the Young Scholars Program of Shandong University (No. 20820162003). GB acknowledges the financial support by the Polish National Science Center through grant UMO-2017/26/D/ST9/01178.

\end{document}